\documentclass[a4paper,11pt]{article}
\usepackage[T1]{fontenc}
\usepackage{lmodern}
\usepackage{microtype}
\usepackage{amsmath,amssymb,mathtools,mathrsfs,booktabs,array}
\usepackage{xcolor}
\usepackage{tikz}
\usepackage{jheppub}

\usepackage[draft,authormarkup=none,commentmarkup=footnote]{changes}
\colorlet{changeadded}{blue!55!white}
\colorlet{changedeleted}{red!55!white}
\setaddedmarkup{\textcolor{changeadded}{#1}}
\setdeletedmarkup{\textcolor{changedeleted}{\ifmmode\text{\sout{\ensuremath{#1}}}\else\sout{#1}\fi}}

\newcommand{\leg}[1]{\mathbf{#1}}
\newcommand{\SU}{\mathrm{SU}}
\newcommand{\Res}{\operatorname*{Res}}
\newcommand{\dd}{\mathrm d}
\newcommand{\B}{B}
\newcommand{\Bg}{B^{\mathrm g}}
\newcommand{\Ah}{\widehat A}
\newcommand{\PP}{\mathscr{P}}

\allowdisplaybreaks[1]

\title{Infrared Consistency and the Uniqueness of String Amplitudes}
\author[c,d,e]{Yu-tin Huang,}
\author[a]{Shi-Lin Wan,}
\author[a]{Zhuo-Hui Wang}
\author[a,b]{and Shuang-Yong Zhou}

\affiliation[a]{Interdisciplinary Center for Theoretical Study, University of Science and Technology of China, Hefei, Anhui 230026}
\affiliation[b]{Peng Huanwu Center for Fundamental Theory, Hefei, Anhui 230026}

\affiliation[c]{Department of Physics and Center for Theoretical Physics, National Taiwan University, Taipei 10617}
\affiliation[d]{Physics Division, National Center for Theoretical Sciences, Taipei 10617}
\affiliation[e]{Max Planck{-}IAS{-}NTU Center for Particle Physics, Cosmology and Geometry, Taipei 10617}

\emailAdd{yutin@phys.ntu.edu.tw}
\emailAdd{wsl9868@mail.ustc.edu.cn}
\emailAdd{wzh33@mail.ustc.edu.cn}
\emailAdd{zhoushy@ustc.edu.cn}

\abstract{
Recent work has shown that maximal supersymmetry, combined with a particular scalar-parity condition at six points, imposes nonlinear constraints on four-point effective field theory amplitudes. Numerical bootstrap studies incorporating unitarity provide evidence for the Veneziano amplitude in maximally supersymmetric Yang--Mills theory and for the Virasoro--Shapiro amplitude, together with infinite-spin alternatives, in supergravity.
In this paper, we establish these constraints to all orders in the low-energy expansion. The derivation uses a collinear limit in which three three-particle factorization channels, or six in gravity, simultaneously go on shell. A scalar-parity projection and a single discrete R-symmetry eliminate the unknown six-point remainder, leaving three-term master equations determined entirely by four-point factorization.
Solving these equations proves the conjectured exponential forms and provides a map that reconstructs the full amplitude from its regular forward limit. Under explicit analytic and positive dispersive assumptions, the fixed reflection-even part of the gauge-theory exponential determines the forward spectral measure and uniquely selects the Veneziano amplitude, with finite-spin residues emerging as a consequence. In gravity, the exponential leaves the forward measure undetermined: a measure supported at a single massive level reconstructs an amplitude with infinitely many exchanged spins at that mass. Finite-spin support at the lowest massive pole is therefore an additional assumption required to select the Virasoro--Shapiro amplitude within the stated ultraviolet framework.
}

\begin{document}
\maketitle
\section{Introduction}

The S-matrix bootstrap seeks to determine how strongly basic principles constrain low-energy interactions and their possible UV completions. Recent developments in positivity bounds have turned analyticity, unitarity and crossing symmetry into increasingly restrictive inequalities on effective field theory (EFT) Wilson coefficients, extending forward-limit positivity to nonforward constraints and two-sided bounds~\cite{Adams:2006sv,deRham:2017avq,Tolley:2020gtv,Caron-Huot:2020cmc} (see \cite{deRham:2022hpx} for a review). In particular, the moment approach interprets dispersive sum rules as moments of positive spectral measures, providing a framework for deriving analytic bounds and reconstructing UV information from low-energy data~\cite{Arkani-Hamed:2020blm,Bellazzini:2020cot,Chiang:2021ziz,Wan:2024eto}. However, current S-matrix bootstrap studies rarely use higher-point amplitudes (see \cite{Cheung:2025nhw,Basile:2026gnd,Cheung:2026lpv,Kremminger:2026edr,Jeong:2026xzk,Backus:2026qdp} for recent developments in this direction), whose consistency conditions provide important additional input for uniquely identifying theories.

String amplitudes offer a natural setting in which to ask whether positivity constraints and additional consistency conditions can determine an entire amplitude. In the absence of supersymmetric protection, the consistency of various deformations of string amplitudes has been examined over the years~\cite{Caron-Huot:2016icg,Sever:2017ylk,Figueroa:2022onw,Huang:2022mdb,Maldacena:2022ckr,Cheung:2022mkw,Bhardwaj:2022lbz,Chen:2023dcx,Cheung:2023adk,Jepsen:2023sia,Cheung:2023uwn,Geiser:2023qqq,Bhardwaj:2024klc,Gadde:2025fil,Cheung:2025nhw}. Numerical studies have shown that positivity bounds supplemented by the string monodromy condition or the higher-point splitting condition confine low-energy coefficients to small regions around the Veneziano values~\cite{Huang:2020nqy,Chiang:2023quf,Berman:2023jys, Berman:2025owb}. Complementary approaches have explored the rigidity of string amplitudes through multiparticle factorization~\cite{Arkani-Hamed:2023jwn}, high-energy softness together with level truncation or minimal zeros~\cite{Cheung:2024uhn,Cheung:2025tbr}, minimal entanglement \cite{Bhat:2024agd}, bootstrap locations of string amplitudes \cite{Guerrieri:2021ivu, Albert:2024yap, Boisvert:2026sfh}, and consistency conditions on generalized Veneziano and Virasoro--Shapiro amplitudes~\cite{Geiser:2022exp,Cheung:2024obl}. More recently, the moment approach has led to analytic uniqueness proofs for the Veneziano amplitude, using either monodromy or splitting and hidden-zero conditions together with the corresponding dispersive and spectral assumptions~\cite{Wan:2026pjq}. These developments motivate identifying low-energy principles from which the additional structure used in the string bootstrap can itself be derived. For example, locality together with the KKBCJ/double-copy construction can give rise to string monodromy in a low-energy EFT setup~\cite{Chen:2022shl}.

Our focus here is on maximal supersymmetry. Its implications depend crucially on the number of external particles. At four points, the supersymmetry Ward identities fix the dependence on external states while leaving substantial freedom in the higher-derivative interactions~\cite{Elvang:2009wd,Elvang:2013cua,Guerrieri:2022sod,Guerrieri:2021ivu,Berman:2024wyt,Berman:2024eid}. At six points, tree-level factorization introduces products of four-point amplitudes, and compatibility with supersymmetry can therefore constrain the four-point coefficients nonlinearly. The constraining power of this approach was explored in Ref~\cite{Elvang:2026pmc}: for planar $\mathcal N=4$ Yang--Mills EFTs, supersymmetry, SU(4) R-symmetry, massless factorization and a scalar-parity condition yield restrictive relations among the four-point Wilson coefficients. A related analysis of $\mathcal N=8$ supergravity EFTs, with SU(4)$\times$ SU(4) R-symmetry and a corresponding scalar-parity condition, revealed analogous relations in gravity~\cite{Berman:2026ezk}. 
Incorporating these order-by-order results into the positivity bootstrap again constrains the low-order coefficients to small regions around the string values. These results also led to conjectured exponential ans\"atze for the four-point amplitudes, with the remaining freedom encoded only in forward coefficients~\cite{Elvang:2026pmc, Berman:2026ezk}. With the exponential ansatz, one can use the analytic bootstrap method of \cite{Wan:2026pjq} to uniquely select the Virasoro--Shapiro amplitudes from the SUGRA EFT \cite{Berman:2026ezk}. These results provide strong evidence for string uniqueness under the stated assumptions, but an all-order derivation of the nonlinear constraints and the conjectured exponential forms is needed to complete the argument.

In this work, we establish these constraints to all orders in the low-energy expansion for both Yang--Mills and supergravity EFTs. Our approach organizes six-point consistency into closed master equations for the four-point amplitudes and makes the role of each assumption explicit. For maximal super Yang-Mills, with 
\begin{equation}
\mathcal{A}_4=-\frac{\Delta_4}{su}B(s,u)\,,
\end{equation}
where $\Delta_4$ is the supersymmetric delta function required by super Ward identities. 
We find the following quadratic master equation for the reduced scalar function $B(s,u)$ 
\begin{align}
\boxed{\begin{aligned}
&~~~(x_1x_3-x_2x_4)\B(x_1,x_2)\B(x_3,x_4) \\
&+ x_2x_4\B(x_1,x_2+x_3)\B(x_3,x_1+x_4) \\
&=x_1x_3\B(x_2,x_1+x_4)\B(x_2+x_3,x_4) ~ .
\end{aligned}}
\end{align}
where $x_1, x_2, x_3$ are free variables and $\sum_{i=1}^4 x_i=0$. For maximal supergravity, 
\begin{equation}
\mathcal{A}^g_4=\frac{\kappa^2\Delta_4^{\rm L}\Delta_4^{\rm R}}{stu}B^{\mathrm g}(s,u)\,,
\end{equation}
the master equation takes the form, 
\begin{align}
\boxed{\begin{aligned}
0={}&(x_1-x_2)\Bg(x_1,x_2)\Bg(x_1+x_2,x_3)\\
&+(x_2+x_3)\Bg(x_2,x_3)\Bg(x_1,x_2+x_3)\\
&-(x_1+x_3)\Bg(x_1,x_3)\Bg(x_2,x_1+x_3)\,.
\end{aligned}}
\end{align}
The key is a scalar-parity projection in a special collinear limit where two external momenta coincide and several factorization channels simultaneously go on shell. Supersymmetry and an appropriate R-symmetry make the unknown six-point remainder drop out of this projection, while the pole terms leave finite contributions determined entirely by products of four-point amplitudes. 
In particular, we show that invariance under a single discrete R-symmetry transformation suffices for the argument, weakening the continuous R-symmetry assumptions of the earlier analyses.

Solving the master equations proves the previously conjectured exponential
forms and clarifies their relation to other structures used in the string
bootstrap. In the gauge theory, the master equation implies the four-point
splitting identity and string monodromy, and fixes the reflection-even part
of the exponential generator in terms of the leading higher-derivative
coupling. These are infrared results obtained before imposing ultraviolet
consistency conditions. Under positive forward dispersion, the fixed even
part determines the full spectral measure by moment determinacy, following
Ref.~\cite{Wan:2026pjq}, and reconstructs the Veneziano amplitude. In gravity the generator is odd but its forward coefficients remain free.
The same reconstruction procedure, applied to a measure supported at one
massive level, produces a rational amplitude whose residue requires
infinitely many spins at that mass. After deriving its off-forward
form and spin content we identify it with the infinite spin tower of
Ref.~\cite{Berman:2026ezk}. Finite-spin support at the lowest massive pole
then supplies an additional selection condition in the gravitational
uniqueness argument. Thus the different roles of finite-spin support follow
from the different information contained in the two exponential solutions.

The paper is organized as follows. In Section~\ref{sec:sym}, we derive the gauge-theory master equation from six-point supersymmetry, tree-level factorization and scalar parity, and show that a single discrete R-symmetry suffices to eliminate the unknown six-point contributions. Section~\ref{sec:sym-solution} solves this equation to all orders in the low-energy expansion, establishing the exponential form, and shows how positive forward dispersive sum rules fix the remaining freedom and uniquely select the Veneziano amplitude. Section~\ref{sec:gravity} extends the framework to $\mathcal N=8$ supergravity EFTs and derives the gravitational master equation and its general exponential solution. Section~\ref{sec:discussion} summarizes these results and discusses open questions. The appendices provide details of the scalar projections, local regularity arguments, residue calculations, and derivation of string monodromy.

\paragraph{Note added:}
During the completion of this draft, the work of Sung \cite{Sung:2026euf} appeared on arXiv.
It takes a different two-channel collinear limit to analyze the six-point factorization for SYM, and derives the unrestricted splitting identity~\eqref{eq:splittingCon0} that is a corollary of our three-term equation \eqref{eq:sym-master}. The three-term master equation allows us to rigorously obtain the stronger exponential solution \eqref{eq:exp_ansatz}, and can be readily generalized to the SUGRA case.  

\section{Maximal SUSY, six-point factorization and scalar parity}
\label{sec:sym}

In this section, we develop the framework for planar $\mathcal N=4$ Yang--Mills effective field theories.
Six-point factorization determines the residues in the three-particle channels from four-point amplitudes, but leaves independent regular six-point contributions undetermined.
Our key observation is that, after subtracting supersymmetric pole terms, these unknown contributions can be eliminated in a special limit where two external momenta coincide and the three-particle channels simultaneously go on shell.
Supersymmetry and a single discrete R-symmetry ensure that the two scalar components related by scalar parity agree on the regular remainder at that locus.
The pole terms retain finite, nonzero contributions to their difference, with kinematic factors multiplying products of four-point amplitudes.
Imposing scalar parity on the full six-point amplitude therefore yields a closed master equation relating three such products.

In Section~\ref{sec:superamplitude}, we describe the four- and six-point superamplitudes in non-chiral superspace.
Section~\ref{sec:six-point} constructs the supersymmetric pole terms from tree-level factorization.
In Section~\ref{sec:scalar-parity}, we establish the cancellation of the regular remainder under the scalar-parity projection in the collinear limit and identify the discrete R-symmetry sufficient for it.
Section~\ref{sec:sym-consistency-equation} then combines these ingredients to obtain the master equation, which will be solved to all orders in Section~\ref{sec:sym-solution}.

\subsection{\texorpdfstring{$\mathcal N=4$}{N=4} superamplitudes}
\label{sec:superamplitude}

Consider tree amplitudes of a single massless $\mathcal N=4$ vector multiplet in four dimensions, containing a gluon, four gluinos and six real scalars.
We work with color-ordered amplitudes at a fixed cyclic ordering, so planar propagators carry sums of consecutive external momenta.
At each EFT order, we allow all local contact terms consistent with the stated symmetries, together with standard simple massless propagator poles.
The three-point amplitudes are those of Yang--Mills theory, and only this multiplet is exchanged at massless poles.
We initially solve the supersymmetry constraints without imposing R-symmetry on the six-point amplitude.
The discrete R-symmetry and the separate scalar-parity assumption will be specified in Section~\ref{sec:scalar-parity}.

We use spinor-helicity variables to describe the external states. A null momentum is represented by a rank-1 bispinor $p_{\leg i}=\lambda_{\leg i}\widetilde\lambda_{\leg i}$, where angle and square brackets denote antisymmetric contractions of the two-component spinors. We allow the spinors to be complex, so that $\lambda$ and $\widetilde\lambda$ are independent. This lets us approach several factorization poles at once. With all momenta outgoing and external-leg labels in bold\footnote{The boldfaced indices denote external massless particles for readability and should not be confused with the commonly used convention for massive particles.}, our conventions are
\begin{align}
&p_{\leg i}=\lambda_{\leg i}\widetilde\lambda_{\leg i},
~~~\sum_{\leg i}p_{\leg i}=0,
~~~s_{\leg{ij}}=\langle\leg{ij}\rangle[\leg{ij}]=(p_{\leg{i}}+p_{\leg{j}})^2, \nonumber \\
&~~~~~~~s_{\leg{ijk}}=s_{\leg{ij}}+s_{\leg{ik}}+s_{\leg{jk}}=(p_{\leg{i}}+p_{\leg{j}}+p_{\leg{k}})^2.
\label{eq:kin}
\end{align}
For a cyclic four-point ordering, we also use the standard Mandelstam variables $(s,t,u)=(s_{\leg{12}},s_{\leg{13}},s_{\leg{14}})$, with $s+t+u=0$.
At four points, maximal supersymmetry fixes the state dependence up to a single scalar function $A(s,u)$.
Normalizing the nonzero Yang--Mills coupling to one, we define
\begin{align}
\mathcal A_4[--++]&=\langle\leg{12}\rangle^2[\leg{34}]^2A(s,u),& A(s,u)&=-\frac1{su}+\Ah(s,u),\nonumber\\
\B(s,u)&=-suA(s,u),& \Ah(s,u)&=\sum_{n\geq0}\sum_{q=0}^{n}a_{n,q}s^{n-q}u^q.
\label{eq:four}
\end{align}
Here $\mathcal A_n$ denotes a superamplitude, a generating super-polynomial whose coefficients are amplitudes for specified external states.
The brackets select a component; in the first line, $--++$ specifies the four gluon helicities.
The function $\Ah$ contains the local four-point corrections, while $\B$ is the full four-point function divided by its Yang--Mills value.
Cyclicity gives $\B(s,u)=\B(u,s)$ and hence $a_{n,q}=a_{n,n-q}$; the normalization also implies $\B(s,0)=1$.
All low-energy expansions below are understood order by order, without assuming convergence of the infinite series.

\paragraph{Non-chiral superspace.}
For our calculation, we use a non-chiral representation that treats the two spinor chiralities on equal footing.

We split the four chiral flavor labels into two pairs and Fourier transform the second pair.
The notation is adapted to an $\SU(2)\times \SU(2)$ subgroup of $\SU(4)$, but this choice of coordinates does not impose invariance under that subgroup.  
The Grassmann variables are denoted by
\begin{equation}
    \eta_{\leg i}^I,\qquad \hat\eta_{\leg i I'},\qquad I,I'=1,2.
\end{equation}
The corresponding supercharges are
\begin{align}
Q^{\alpha I} &= \sum_{\leg i} \lambda_{\leg i}^\alpha \eta_{\leg i}^I,
&
\bar Q^{\dot\alpha}{}_{I}
&=\sum_{\leg i} \tilde\lambda_{\leg i}^{\dot\alpha}
\frac{\partial}{\partial \eta_{\leg i}^I},
\\
\hat Q^{\dot\alpha}{}_{I'}
&=\sum_{\leg i} \tilde\lambda_{\leg i}^{\dot\alpha}\hat\eta_{\leg i I'},
&
\hat{\bar Q}^{\alpha I'}
&=\sum_{\leg i} \lambda_{\leg i}^\alpha
\frac{\partial}{\partial\hat\eta_{\leg i I'}}.
\end{align}
The variables $\eta$ and $\hat\eta$ carry helicities
$+\frac12$ and $-\frac12$, respectively.  Hence the non-chiral on-shell
superfield is helicity neutral,
\begin{equation}
\Phi(\eta,\hat\eta)
=
\phi
+\eta^I\psi_I^{-1/2}
+\hat\eta_{I'}\psi^{I',+1/2}
+\eta^2 g^-
+\hat\eta^2 g^+
+\cdots
+\eta^2\hat\eta^2\phi''.
\end{equation}
Consequently a physical superamplitude must have vanishing total helicity
weight on every leg. One can obtain the non-chiral amplitudes from chiral ones via simple Fourier transform:
\begin{align}
\mathcal A_{\mathrm{nonchiral}}
=\left[\prod_{\leg i}\int\!\dd\eta_{\leg i}^4\,\dd\eta_{\leg i}^3\,
 e^{\eta_{\leg i}^3\hat\eta_{\leg i1'}+\eta_{\leg i}^4\hat\eta_{\leg i2'}}\right]
 \mathcal A_{\mathrm{chiral}}\,.
\end{align}

\paragraph{Solving SUSY Ward identities in the selected sector.}
We need only the flavor-occupation sector contributing to the two six-scalar components below.
Each of the four chiral flavors occurs three times in these components; the half-Fourier transform preserves this occupation number at six points.
After factoring out the multiplicative supercharges, write
\begin{align}
\mathcal A_{6,\mathrm{sel}}=\Delta_6\mathcal P^{\rm NMHV}_6,
\qquad \Delta_6=\delta^4(Q)\delta^4(\hat Q).
\label{eq:charge-factor}
\end{align}
The selected reduced polynomial has one residual Grassmann variable from each flavor family. The derivative supercharges require the coefficient vectors of these linear forms to obey
\begin{align}
\sum_{\leg i}\widetilde\lambda_{\leg i}^{\dot\alpha}\widetilde\nu_{\leg i}=0,
\qquad \sum_{\leg i}\lambda_{\leg i}^\alpha\nu_{\leg i}=0.
\label{eq:ward-kernels}
\end{align}
Momentum conservation gives $\operatorname{row}\lambda\subset\ker\widetilde\lambda$ and $\operatorname{row}\widetilde\lambda\subset\ker\lambda$. Shifts by these charge directions are annihilated by $\Delta_6$, so the relevant spaces are the quotients
\begin{align}
E=\ker\widetilde\lambda/\operatorname{row}\lambda,
\qquad \widehat E=\ker\lambda/\operatorname{row}\widetilde\lambda,
\qquad \dim E=\dim\widehat E=2.
\label{eq:residual-spaces}
\end{align}
Choose locally regular dual representatives $\widetilde\nu^a,\nu_a$ with $\widetilde\nu^a\cdot\nu_b=\delta^a{}_b$, and define
\begin{align}
\theta^{aI}=\sum_{\leg i}(\widetilde\nu^a)_{\leg i}\eta_{\leg i}^I,
\qquad \hat\theta_{aI'}=\sum_{\leg i}(\nu_a)_{\leg i}\hat\eta_{\leg i I'}.
\end{align}
The general solution in the selected sector is then
\begin{align}
\mathcal P^{\rm NMHV}_6
&=C_{ab}{}^{cd}\,\theta^{a1}\theta^{b2}\hat\theta_{c1'}\hat\theta_{d2'},
\qquad a,b,c,d=1,2.
\label{eq:selected-ward}
\end{align}
The sixteen coefficients are independent before an R-symmetry is imposed.
Other flavor occupations can occur in a superamplitude without continuous R-symmetry, but do not contribute to the scalar projections used here.
Throughout the derivation, $\mathcal P^{\rm NMHV}_6$ denotes the selected flavor-occupation sector relevant to these scalar projections.
A locally nonsingular basis, needed to transfer regularity from components to these coefficients, is justified in Appendix~\ref{app:local-regularity}.

\subsection{Six-point factorization and pole subtraction}
\label{sec:six-point}

\begin{figure}[t]
\centering
\begingroup
\definecolor{symSideA}{HTML}{168AC2}
\definecolor{symSideB}{HTML}{E58324}
\begin{tikzpicture}[x=0.9cm,y=0.9cm,
  sym order/.style={draw=black!40,line width=0.45pt,
    dash pattern=on 2.2pt off 2.2pt},
  sym edge A/.style={draw=symSideA,line width=0.75pt},
  sym edge B/.style={draw=symSideB,line width=0.75pt},
  sym propagator/.style={draw=black!85,line width=0.85pt,
    preaction={draw=white,line width=2.8pt}},
  sym blob A/.style={circle,draw=symSideA,fill=symSideA!8,
    line width=0.7pt,minimum size=7.8mm,inner sep=0.8pt,font=\footnotesize},
  sym blob B/.style={circle,draw=symSideB,fill=symSideB!8,
    line width=0.7pt,minimum size=7.8mm,inner sep=0.8pt,font=\footnotesize},
  sym leg/.style={font=\footnotesize,inner sep=1pt,fill=white},
  sym title/.style={font=\footnotesize,inner sep=0pt}]
\path[use as bounding box] (-2.02,-2.20) rectangle (10.82,1.78);
\newcommand{\planarPanel}[6]{
  \begin{scope}[shift={#1}]
    \draw[sym order] (0,0) circle[radius=1.60];
    \foreach \symLeg/\symAngle in {
      1/120,2/180,3/240,4/300,5/0,6/60}{
      \coordinate (symE\symLeg) at (\symAngle:1.60);
      \coordinate (symL\symLeg) at (\symAngle:1.84);
    }
    \coordinate (symA) at (#4:0.64);
    \coordinate (symB) at ({#4+180}:0.64);
    \foreach \symLeg in {#6}{
      \draw[sym edge B] (symB)--(symE\symLeg);
    }
    \foreach \symLeg in {#5}{
      \draw[sym edge A] (symA)--(symE\symLeg);
    }
    \draw[sym propagator] (symA)--(symB);
    \node[sym blob A] at (symA) {$\mathcal A_4$};
    \node[sym blob B] at (symB) {$\mathcal A_4$};
    \foreach \symLeg in {1,...,6}{
      \node[sym leg] at (symL\symLeg) {$\mathbf{\symLeg}$};
    }
    \node[sym title] at (0,-1.98)
      {\textbf{(#2)}\quad $s_{\mathbf{#3}}=0$};
  \end{scope}
}
\planarPanel{(0,0)}{a}{123}{180}{1,2,3}{4,5,6}
\planarPanel{(4.40,0)}{b}{234}{240}{2,3,4}{5,6,1}
\planarPanel{(8.80,0)}{c}{345}{300}{3,4,5}{6,1,2}
\end{tikzpicture}
\endgroup
\caption{The three planar four-point factorizations at six points.
Each blob is a full on-shell four-point superamplitude, and each internal line sums over the massless $\mathcal N=4$ multiplet.
Continuing the residues away from their cuts defines the pole terms in \eqref{eq:pole-decomposition}.}
\label{fig:six-factorization}
\end{figure}
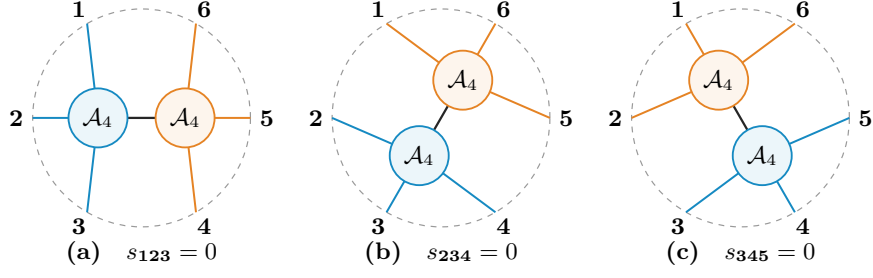

The Ward identities determine the allowed state dependence, but leave the functions $C_{ab}{}^{cd}$ unfixed.
Factorization fixes their residues at three-particle poles.
For the ordering $(\leg1,\ldots,\leg6)$, the distinct planar three-particle channels are $\leg{123}$, $\leg{234}$ and $\leg{345}$.
In each channel, we multiply two full four-point superamplitudes and sum over the states of the internal multiplet, as illustrated in Figure~\ref{fig:six-factorization}.
This operation, called sewing, separates the residue into a supersymmetric numerator and a product of the state-independent functions $A$ in \eqref{eq:four}.

\paragraph{The on-shell residue.}
The four-point non-chiral superamplitude is $\mathcal A_4=\Delta_4A(s,u)$ in our conventions.
The factorization pole, $s_{\leg T}=0$ puts the internal momentum on shell. Here $\leg T=(\leg a\leg b\leg c)$ represents an arbitrary 3-tuple, and $\leg T^c$ its complement. On the pole, $p_\ell=\sum_{\leg i\in\leg T}p_{\leg i}=\lambda_\ell\widetilde\lambda_\ell$ and we use outgoing internal momenta $-p_\ell$ and $p_\ell$ on the left and right.
We choose $\lambda_{-\ell}=-\lambda_\ell$, $\widetilde\lambda_{-\ell}=\widetilde\lambda_\ell$, along with $\eta_{-\ell}=\eta_\ell$ and $\hat\eta_{-\ell}=-\hat\eta_\ell$.
Writing $q_{\leg T}$ and $\hat q_{\leg T}$ for the charges summed over the triple, the two delta factors are
\begin{align}
\Delta_{\mathrm L}&=\delta^4(q_{\leg T}-\lambda_\ell\eta_\ell)\delta^4(\hat q_{\leg T}-\widetilde\lambda_\ell\hat\eta_\ell),&\Delta_{\mathrm R}&=\delta^4(q_{\leg T^c}+\lambda_\ell\eta_\ell)\delta^4(\hat q_{\leg T^c}+\widetilde\lambda_\ell\hat\eta_\ell).
\end{align}
The residue for $1/s_{\leg T}$ is given by the product of two four-point amplitudes. For example, for $\leg T=\leg{123}$, the four-point function is of momentum  $(\leg1,\leg2,\leg3,-\ell)$ and $(\leg4,\leg5,\leg6,\ell)$, with
\begin{align}
(s_{\mathrm L},u_{\mathrm L})=(s_{\leg{12}},s_{\leg{23}}),\qquad (s_{\mathrm R},u_{\mathrm R})=(s_{\leg{45}},s_{\leg{56}}).
\end{align}
We use the following internal-state sewing convention, with the common overall sign fixed consistently with the pole numerator below.
This common sign cancels from the homogeneous master equation: 
\begin{align}
\Res_{s_{\leg T}=0}\mathcal A_{6,\mathrm{sel}}&=-\int\!\dd\hat\eta_{\ell2'}\,\dd\hat\eta_{\ell1'}\,\dd\eta_\ell^2\,\dd\eta_\ell^1\,\Delta_{\mathrm L}\Delta_{\mathrm R}A(s_{\mathrm L},u_{\mathrm L})A(s_{\mathrm R},u_{\mathrm R})\nonumber\\
=&-\Delta_6\langle\ell q_{\leg T}^{1}\rangle\langle\ell q_{\leg T}^{2}\rangle[\ell\hat q_{\leg T,1'}][\ell\hat q_{\leg T,2'}]A(s_{\mathrm L},u_{\mathrm L})A(s_{\mathrm R},u_{\mathrm R}).
\end{align}

To express the result using only external spinors, let us define the following invariants:
\begin{align}
\Gamma_{{\leg T},I'}(\hat{\eta})&=\langle\leg{ab}\rangle \hat{\eta}_{{\leg c},I'}+\langle\leg{bc}\rangle \hat{\eta}_{{\leg a},I'}+\langle\leg{ca}\rangle \hat{\eta}_{{\leg b},I'},\nonumber\\
\widehat\Gamma_{\leg T}^I(\eta)&=[\leg{ab}]\eta^I_{\leg c}+[\leg{bc}]\eta^I_{\leg a}+[\leg{ca}]\eta^I_{\leg b}\,.
\label{eq:minors}
\end{align}
These linear forms vanish under the action of $\hat{\bar{Q}}$ and $\bar{Q}$ respectively. On the pole, we can use the following identities:
\begin{align}
\langle\ell q_{\leg T}^{I}\rangle&=\frac{\langle\ell\leg a\rangle}{[\leg{bc}]}\hat{\Gamma}_{\leg T}^{I},& [\ell\hat q_{\leg T,I'}]&=\frac{[\ell\leg a]}{\langle\leg{bc}\rangle} \Gamma_{\leg T,I'},&\frac{\langle\ell\leg a\rangle[\ell\leg a]}{\langle\leg{bc}\rangle[\leg{bc}]}&=\frac{s_{\leg{ab}}+s_{\leg{ac}}}{s_{\leg{bc}}}=-1.
\end{align}
Defining, 
\begin{align}
\mathcal U_{\leg T}=\hat{\Gamma}_{\leg T}^{1}\hat{\Gamma}_{\leg T}^{2}\Gamma_{\leg T,1'}\Gamma_{\leg T,2'},
\end{align}
the three-particle residue of the six-point NMHV amplitude can be written as,
\begin{align}
\Res_{s_{\leg T}=0}\mathcal A_{6,\mathrm{sel}}=-\Delta_6\mathcal U_{\leg T}A(s_{\mathrm L},u_{\mathrm L})A(s_{\mathrm R},u_{\mathrm R})\equiv \widetilde{\mathcal R}_{\leg T}.
\end{align}

\paragraph{Continuing the residue away from the cut.}
Separating the three-particle poles from the remainder we can write:
\begin{align}
\mathcal A_{6}^{\rm NMHV}=\sum_{\leg T=\leg{123},\leg{234},\leg{345}}\frac{\widetilde{\mathcal R}_{\leg T}}{s_{\leg T}}+\mathcal A_{6,\mathrm{reg}},\qquad \mathcal A_{6,\mathrm{reg}}=\Delta_6\mathcal P^{\rm NMHV}_{6,\mathrm{reg}}\,,
\label{eq:pole-decomposition}
\end{align}
where $\mathcal A_{6,\mathrm{reg}}$ is regular in three-particle poles, but retains two-particle poles.
Introducing
\begin{align}
\Pi_{\leg{123}}&=\B(s_{\leg{12}},s_{\leg{23}})\B(s_{\leg{45}},s_{\leg{56}}),\nonumber\\
\Pi_{\leg{234}}&=\B(s_{\leg{23}},s_{\leg{34}})\B(s_{\leg{16}},s_{\leg{56}}),\nonumber\\
\Pi_{\leg{345}}&=\B(s_{\leg{34}},s_{\leg{45}})\B(s_{\leg{12}},s_{\leg{16}}), 
\label{eq:sym-products}
\end{align}
for $\leg T=(\leg a\leg b\leg c)$ with cyclic complement $(\leg d\leg e\leg f)$, the $\Delta_6$ reduced decomposition becomes
\begin{align}
\mathcal P_6=-\sum_{\leg T=\leg{123},\leg{234},\leg{345}}\frac{\mathcal U_{\leg T}\Pi_{\leg T}}{s_{\leg T}s_{\leg{ab}}s_{\leg{bc}}s_{\leg{de}}s_{\leg{ef}}}+\mathcal P_{6,\mathrm{reg}}\,.
\end{align}
All undetermined information in the selected six-point sector is now contained in $\mathcal P_{6,\mathrm{reg}}$.
Here and below the external particles remain on shell: continuation away from the cut means varying the internal invariant $s_{\leg T}$, not taking external momenta off shell.

At this point the next step is clear. We would like to devise a projection which imposes a constraint on the first term on the RHS while projecting the remainder out. This is achieved by the collinear limit of scalar parity in the next subsection.

\subsection{Scalar parity and the collinear limit}
\label{sec:scalar-parity}

Let us consider the scalar-parity condition employed  in Ref.~\cite{Elvang:2026pmc},
\begin{align}
A_6[z^{12}_{\leg1}z^{13}_{\leg2}z^{14}_{\leg3}z^{23}_{\leg4}z^{34}_{\leg5}z^{24}_{\leg6}]&=A_6[z^{34}_{\leg1}z^{24}_{\leg2}z^{23}_{\leg3}z^{14}_{\leg4}z^{12}_{\leg5}z^{13}_{\leg6}]\,.
\label{eq:parity-assumption}
\end{align}
The right-hand side replaces every antisymmetric scalar flavor pair by its complement in $\{1,2,3,4\}$, with the state phases fixed by our superspace conventions. We translate this scalar projection to our non-chiral superspace. 
For example, on the LHS, flavor 1 occurs in $z^{12}_{\leg1}$, $z^{13}_{\leg2}$ and $z^{14}_{\leg3}$, so we want to project out the 3-tuple $\leg{123}$, i.e., $\eta^1_1\eta^1_2\eta^1_3$. 
All in all,  chiral flavors $(1,2,3,4)$ occupy the leg triples $(\leg{123},\leg{146},\leg{245},\leg{356})$, respectively. The half-Fourier transform leaves the first two triples unchanged and replaces each of the last two by its complement.
Thus the non-chiral flavor families $(1,2,1',2')$ occupy $(\leg{123},\leg{146},\leg{136},\leg{124})$.
For the RHS of \eqref{eq:parity-assumption}, the corresponding non-chiral triples are $(\leg{456},\leg{235},\leg{245},\leg{356})$. Importantly, all triples contain distinct elements. 

Let $\varphi[\mathcal P^{\rm NMHV}_6]$ and $\overline\varphi[\mathcal P^{\rm NMHV}_6]$ extract the left and right components of \eqref{eq:parity-assumption} from $\Delta_6\mathcal P^{\rm NMHV}_6$.
Defining their difference by, 
\begin{align}
\PP\equiv\overline\varphi-\varphi ,
\end{align}
the scalar-parity then requires $\PP(\mathcal P_6)=0$.
For an ordered triple $\leg i<\leg j<\leg k$, in general, their projection on $\Delta_6 \mathcal{P}^{\rm NMHV}$ can be encapsulated by the identity: 
\begin{align}
\delta^2(Q^I)\biggl(\sum_{\leg l}\widetilde \omega_{\leg l}\eta_{\leg l}^I\biggr)\bigg|_{\eta_{\leg i}^I\eta_{\leg j}^I\eta_{\leg k}^I}
&=\langle\leg{ij}\rangle\widetilde \omega_{\leg k}-\langle\leg{ik}\rangle\widetilde \omega_{\leg j}+\langle\leg{jk}\rangle\widetilde \omega_{\leg i}=\Gamma_{\leg{ijk}}(\widetilde \omega)\,,
\nonumber
\end{align}
where we borrow the notation $\Gamma_{{\leg T}}(\widetilde \omega)$ to represent the bosonic function defined for the triple as opposed to fermionic $\Gamma_{{\leg T},I'}$. 
For the hatted flavors $\hat{\eta}$, we simply replace angle brackets by square brackets and obtain $\widehat\Gamma_{\leg{ijk}}(\omega)$. Thus we find that each of the four triples contributes a factor of $\Gamma$ or $\hat{\Gamma}$. So we have 
\begin{align}
\varphi[\mathcal{P}_6^{\rm NMHV}]&=C_{ab}\,^{cd}\Gamma_{\leg{123}}(\widetilde\nu^a)\Gamma_{\leg{146}}(\widetilde\nu^b)\widehat\Gamma_{\leg{136}}(\nu_c)\widehat\Gamma_{\leg{124}}(\nu_d),\nonumber\\
\overline\varphi[\mathcal{P}_6^{\rm NMHV}]&=C_{ab}\,^{cd}\Gamma_{\leg{456}}(\widetilde\nu^a)\Gamma_{\leg{235}}(\widetilde\nu^b)\widehat\Gamma_{\leg{245}}(\nu_c)\widehat\Gamma_{\leg{356}}(\nu_d)\,.
\label{eq:two-products}
\end{align}

\paragraph{Relating the projections in the collinear limit.}
The two expressions in \eqref{eq:two-products} are generally different.
They become related when two nonzero external momenta coincide and the three planar channels meet:
\begin{align}
p_{\leg6}=p_{\leg3},\qquad s_{\leg{123}}=s_{\leg{234}}=s_{\leg{345}}=0.
\label{eq:collision}
\end{align}
This collinear limit corresponds to setting $\lambda_{\leg6}=\lambda_{\leg3}$ and $\widetilde\lambda_{\leg6}=\widetilde\lambda_{\leg3}$.
Writing $x_1=s_{\leg{23}}$, $x_2=s_{\leg{34}}$ and $x_3=s_{\leg{56}}$, momentum conservation gives
\begin{align}
(s_{\leg{12}},s_{\leg{23}},s_{\leg{34}},s_{\leg{45}},s_{\leg{56}},s_{\leg{16}})=(x_2+x_3,x_1,x_2,-x_2-x_3,x_3,-x_1-x_2-x_3).
\label{eq:adjacent}
\end{align}
The same momenta also obey
\begin{align}
s_{\leg{36}}=s_{\leg{134}}=s_{\leg{135}}=s_{\leg{146}}=0.
\label{eq:additional-zero-channels}
\end{align}
These are consequences, not additional independent conditions on $x_1,x_2,x_3$. We work at generic points where the adjacent invariants and the bracket denominators below are nonzero. An explicit spinor parameterization and the corresponding rank checks are given in Appendix~\ref{app:local-regularity}.

Note that in the collinear limit, the vectors $(\nu, \tilde{\nu})$ satisfy the following kinematic identities
\begin{align}\label{eq:limit-identities}
\langle\leg{12}\rangle\Gamma_{\leg{456}}(\widetilde \nu)&=-\langle\leg{45}\rangle\Gamma_{\leg{123}}(\widetilde \nu),
&\langle\leg{25}\rangle\Gamma_{\leg{146}}(\widetilde \nu)&=\langle\leg{14}\rangle\Gamma_{\leg{235}}(\widetilde \nu),\nonumber\\
[\leg{35}]\widehat\Gamma_{\leg{136}}(\nu)&=[\leg{13}]\widehat\Gamma_{\leg{356}}(\nu), &
\langle\leg{13}\rangle\widehat\Gamma_{\leg{124}}(\nu)&=\langle\leg{35}\rangle\widehat\Gamma_{\leg{245}}(\nu).
\end{align}
Their proof is given in Appendix~\ref{app:scalar-projection-proof}. 
Substituting them into \eqref{eq:two-products}, it is easy to see that it simply equates the projection $\varphi$ with $\bar\varphi$ up to interchanging $c\leftrightarrow d$ and an overall factor $-\frac{\langle12\rangle\langle14\rangle[13]\langle35\rangle}{\langle45\rangle\langle25\rangle[35]\langle13\rangle}$ which evaluates to 1 in our collinear limit, as we show in the Appendix~\ref{app:projection-normalization}.
For coefficients regular at the locus, the two projection maps therefore obey
\begin{align}
\PP(\mathcal P^{\rm NMHV}_6)=\sum_{a,b,c,d=1}^{2}\bigl(C_{ab}{}^{dc}-C_{ab}{}^{cd}\bigr)\Gamma_{\leg{123}}(\widetilde\nu^a)\Gamma_{\leg{146}}(\widetilde\nu^b)\widehat\Gamma_{\leg{136}}(\nu_c)\widehat\Gamma_{\leg{124}}(\nu_d),\quad \text{on }\eqref{eq:collision}.
\label{eq:projection-antisymmetric}
\end{align}
This is an identity on the coincident-momentum locus and requires regular coefficients for direct restriction to that locus.
It selects a kinematic contraction of the part antisymmetric in $c,d$; vanishing of that contraction does not require the whole antisymmetric tensor to vanish.

A hatted $SU(2)$ singlet condition is sufficient to make this projection vanish, but is not equivalent to the selected scalar equality.
In fact, a single discrete transformation already suffices.
Consider the signed flavor exchange
\begin{align}
\mathsf R:\quad(\eta^1,\eta^2,\eta^3,\eta^4)\longmapsto(-\eta^1,\eta^2,\eta^4,\eta^3),\qquad \mathsf R^2=1,\qquad \mathbb Z_2^{\rm R}=\{1,\mathsf R\}.
\end{align}
It is an element of $\SU(4)$: the sign on flavor one compensates the determinant of the exchange.
After the half-Fourier transform, the interchange of the two integration variables contributes one sign per leg, hence an overall $(-1)^6=1$.
The charge factor $\Delta_6$ is invariant, while the residual coordinates and their ordered monomials transform as
\begin{align}
(\theta^{a1},\theta^{a2},\hat\theta_{a1'},\hat\theta_{a2'})&\longmapsto(-\theta^{a1},\theta^{a2},\hat\theta_{a2'},\hat\theta_{a1'}),
\end{align}
Thus, in the selected sector,
\begin{align}
\mathsf R\,\mathcal P^{\rm NMHV}_6=\mathcal P^{\rm NMHV}_6\quad\Longleftrightarrow\quad C_{ab}{}^{cd}=C_{ab}{}^{dc}.
\label{eq:discrete-coefficients}
\end{align}

We henceforth impose this discrete R-symmetry on the full amplitude.
In the selected sector, it leaves twelve invariant structures: four choices of $(a,b)$ and three choices of the symmetric pair $(c,d)$.
Equivalently, the kinematic relation is
\begin{align}
\overline\varphi=\varphi\circ\mathsf R
\quad\text{on the selected Ward space at }\eqref{eq:collision}.
\label{eq:projection-map-identity}
\end{align}
The two assumptions have separate roles.
Discrete R-symmetry makes the scalar difference vanish on invariant regular contributions at this locus.
Physical scalar parity, \eqref{eq:parity-assumption}, equates the full component amplitudes at nearby nonsingular momenta and will constrain the finite terms inherited from their poles.
In particular, for a regular invariant reduced contribution $H$,
\begin{align}
\mathsf RH=H
\quad\Longrightarrow\quad
\left.\PP(H)\right|_{\eqref{eq:collision}}=0.
\label{eq:regular-projection-lemma}
\end{align}
No converse is asserted, and no decomposition of $\PP(H)$ into planar channel invariants alone is needed away from the coincident-momentum surface.

\paragraph{Applying the identity to the subtracted remainder.}
At fixed EFT order, the only planar poles near a generic point of \eqref{eq:collision} are $s_{\leg{123}},s_{\leg{234}},s_{\leg{345}}$.
They can be included in local coordinates, and all adjacent two-particle invariants are nonzero.
Writing each subtracted component over their product, matching the residue on each cut makes the regular numerator divisible by that channel coordinate and hence by their product.
Thus $\mathcal A_{6,\mathrm{reg}}$ is regular also at the intersection.
Both spinor matrices retain rank two, so a locally nonsingular change of Grassmann coordinates factors out $\Delta_6$ without introducing a singular inverse matrix.
Consequently the coefficients of $\mathcal P_{6,\mathrm{reg}}$ in \eqref{eq:selected-ward} are regular; Appendix~\ref{app:local-regularity} supplies the details.

The subtraction preserves the imposed symmetry.
In $\mathcal U_{\leg T}$ the sign on the unhatted flavor-one factor cancels the sign from restoring the order of the two hatted factors, so $\mathsf R\mathcal U_{\leg T}=\mathcal U_{\leg T}$.
The remainder therefore inherits both supersymmetry and $\mathsf R$-invariance, and \eqref{eq:regular-projection-lemma} applies to it.

\paragraph{Finite pole terms and independence of the continuation.}
For a pole term we must project before restricting to the cut.
Sewing the four-point scalar flavor tensors preserves flavor complementation, so the two selected residues agree on the entire channel surface, not merely at its intersection with the other cuts.
Thus $\PP(\mathcal U_{\leg T})=s_{\leg T}K_{\leg T}$ with $K_{\leg T}$ regular, and
\begin{align}
\PP\!\left(\frac{\mathcal U_{\leg T}}{s_{\leg T}}\right)=K_{\leg T}.
\end{align}
The restriction of $K_{\leg T}$ need not vanish.
For example, Appendix~\ref{app:pole-projections} exhibits a factor of $s_{\leg{123}}$ in one projected numerator while its complementary projection vanishes identically.
Equality of residues removes a singularity in the scalar difference, not its finite value.

Two admissible supersymmetric, $\mathsf R$-invariant continuations of the same residue differ by $s_{\leg T}V_{\leg T}$, with $V_{\leg T}$ regular near the locus.
The resulting change in the pole subtraction is regular and has zero projected value by \eqref{eq:regular-projection-lemma}.
The finite constraint is therefore independent of this continuation freedom.
An arbitrary componentwise continuation is not sufficient: the subtraction and remainder must assemble into supersymmetric, symmetry-invariant objects.

\subsection{Master equation}
\label{sec:sym-consistency-equation}
We can now assemble the result. In the collinear limit, the scalar-parity projection will vanish on the three-particle regular part of the amplitude: 
\begin{equation}
\left.\PP(\mathcal P^{\rm NMHV}_{6,\mathrm{reg}})\right|_{\eqref{eq:collision}}=0\,.
\end{equation}
However, the three-particle factorization pole terms require special care because they are singular. We first project with $\PP$ then take the collinear limit. This will lead to $0/0$ cancellations leaving behind finite terms. In particular, one finds 
\begin{align}
\left.\biggl(\frac{\PP(\mathcal U_{\leg{123}})}{s_{\leg{123}}},\frac{\PP(\mathcal U_{\leg{234}})}{s_{\leg{234}}},\frac{\PP(\mathcal U_{\leg{345}})}{s_{\leg{345}}}\biggr)\right|_{\eqref{eq:collision}}=\frac{\epsilon}{2}(x_2+x_3,-x_1-x_2,-x_2-x_3),
\end{align}
where $\epsilon=\epsilon(\leg1,\leg2,\leg3,\leg4)$ uses the convention \eqref{eq:epsilon} and obeys $\epsilon^2=-4x_1x_2x_3(x_1+x_2+x_3)$.

Thus in our collinear limit, the scalar-parity condition becomes\begin{align}
0=\PP(\mathcal P^{\rm NMHV}_{6})=-\sum_{\leg T=\leg{123},\leg{234},\leg{345}}\frac{\PP(\mathcal U_{\leg T})\Pi_{\leg T}}{s_{\leg T}s_{\leg{ab}}s_{\leg{bc}}s_{\leg{de}}s_{\leg{ef}}}.
\end{align}
or 
\begin{align}
(x_1+x_2)(x_2+x_3)\Pi_{\leg{234}}+x_2x_4\Pi_{\leg{123}}-x_1x_3\Pi_{\leg{345}}=0\,,
\end{align}
with  $x_4=-x_1-x_2-x_3$. The products are evaluated at their collinear limit arguments.
Using $(x_1+x_2)(x_2+x_3)=x_1x_3-x_2x_4$, substituting \eqref{eq:sym-products} and applying $\B(s,u)=\B(u,s)$ yields
\begin{align}
\boxed{\begin{aligned}
&~~~(x_1x_3-x_2x_4)\B(x_1,x_2)\B(x_3,x_4) \\
&+ x_2x_4\B(x_1,x_2+x_3)\B(x_3,x_1+x_4) \\
&=x_1x_3\B(x_2,x_1+x_4)\B(x_2+x_3,x_4) ~ .
\end{aligned}}
\label{eq:sym-master}
\end{align}
This is the master equation for the normalized four-point function $\B$.
Its independent variables are $x_1,x_2,x_3$, with $x_4$ fixed by their sum. At each EFT order, the left-hand side of \eqref{eq:sym-master} is polynomial in the independent $x_i$, so its vanishing on an open set implies that it vanishes identically. We may therefore differentiate the master equation or specialize it to configurations such as $x_3=0$ order by order when needed, without requiring the original six-point amplitude to be regular there.

An important feature of the master equation is its triangular structure in the EFT expansion, which organizes the constraints underlying the consistency and bootstrap analysis below. Writing $\B=1+\sum_{n\geq1}\B_n$, where $\B_n$ is homogeneous of total degree $n$, each product in the master equation contributes
\begin{equation}
\bigl[\B(\mathbf u)\B(\mathbf v)\bigr]_n
=\B_n(\mathbf u)+\B_n(\mathbf v)
+\sum_{k=1}^{n-1}\B_k(\mathbf u)\B_{n-k}(\mathbf v),
\end{equation}
where $\mathbf u$ and $\mathbf v$ denote pairs of kinematic arguments and $[\cdots]_n$ extracts the homogeneous term of degree $n$. Substituting this expansion into the master equation and matching homogeneous terms yields a triangular system of recursion relations: the new coefficients at each EFT order enter linearly, while the nonlinear terms involve only lower-order coefficients. These relations need not determine all coefficients uniquely; the freedom remaining after imposing the full hierarchy is subject to the subsequent bootstrap constraints. In the next section, rather than solving these relations order by order, we treat the entire hierarchy at once through functional identities.

\section{Exponential solution and Veneziano uniqueness}
\label{sec:sym-solution}

In this section, we solve the master equation~\eqref{eq:sym-master} subject to $\B(s,u)=\B(u,s)$ and $\B(s,0)=1$, with $A(s,u)=-\B(s,u)/(su)$.
The general solution takes the exponential form
\begin{align}
\boxed{\begin{aligned}
\B(s,u)={}&\sqrt{\frac{\pi\alpha' su\sin\!\bigl(\pi\alpha'(s+u)\bigr)}{(s+u)\sin(\pi\alpha's)\sin(\pi\alpha'u)}}\\
&\times\exp\!\left[\sum_{k\geq1}\frac{a_{2k-1,0}}{2k+1}
\bigl(s^{2k+1}+u^{2k+1}-(s+u)^{2k+1}\bigr)\right].
\end{aligned}}
\label{eq:full-ansatz}
\end{align}
The classification in Section~\ref{sec:alg-exponential} is formal and does not require convergence or a real choice of $\alpha'$.
All reciprocals, logarithms and square roots are normalized by their constant term.
In Section~\ref{sec:uniqueness}, we separately impose analyticity near the origin and exact positive forward dispersive sum rules, and show that the remaining odd coefficients are fixed to the Veneziano values, including the Yang--Mills limit when the positive measure vanishes.

\subsection{Exponential ansatz}
\label{sec:alg-exponential}

\paragraph{Reflected master equation.}
The desired exponential form inspires us to separate the parts of $\log\B$ that are even and odd under reflection:
\begin{align}
\B_{\mathrm{even}}(s,u)=\sqrt{\B(s,u)\B(-s,-u)},\qquad
\B_{\mathrm{odd}}(s,u)=\sqrt{\frac{\B(s,u)}{\B(-s,-u)}}.
\label{eq:even-odd}
\end{align}
Both factors are symmetric in their arguments and equal one on either coordinate axis.
$\B(s,u)=\B(u,s)$ also fixes the odd factor on opposite arguments:
\begin{align}
\B_{\mathrm{odd}}(s,-s)=1,\qquad \B(s,-s)=\B_{\mathrm{even}}(s,-s).
\end{align}
For $x_4=-x_1-x_2-x_3$, we collect the weighted terms of the master equation as $X_1,X_2,X_3$ and denote $X_i^{\vee}(x_1,x_2,x_3,x_4)=X_i(-x_1,-x_2,-x_3,-x_4)$.
The original and reflected master equations read
\begin{align}
X_1+X_2=X_3,\qquad X_1^{\vee}+X_2^{\vee}=X_3^{\vee}.
\end{align}

\paragraph{Cocycle equation.}

At $(x_1,x_2,x_3,x_4)=(s,-s,s+u,-s-u)$, the first weight vanishes, giving $\B(s,u)\B(s+u,-u)=\B(-s,-u)\B(u,-s-u)$.
Dividing by the reflected factors and taking normalized square roots yields $\B_{\mathrm{odd}}(s,u)=\B_{\mathrm{odd}}(u,t)=\B_{\mathrm{odd}}(t,s),~s+t+u=0$.
This cyclic identity implies $X_2/X_2^\vee=X_3/X_3^\vee$.
The master equation and its reflection therefore give
\begin{align}
\frac{X_1}{X_1^\vee}=\frac{X_2}{X_2^\vee}=\frac{X_3}{X_3^\vee}.
\end{align}
Cancelling the weights, taking normalized square roots, and applying the cyclic identity at $(x_1,x_2,x_3,x_4)=(s,u,t,-s-t-u)$, we obtain
\begin{align}
\B_{\mathrm{odd}}(s,u)\B_{\mathrm{odd}}(s+u,t)
=\B_{\mathrm{odd}}(u,t)\B_{\mathrm{odd}}(s,t+u).
\end{align}
The common polynomial weights are cancelled before specializing the kinematics.
The resulting formal power-series identities remain valid under the specializations used below, even when a weight vanishes.

After cancelling the common odd product, the even factor satisfies the same master equation.
Its specialization to $(s,u,-s,-u)$ reads
\begin{align}
(u^2-s^2)\B_{\mathrm{even}}(s,u)^2
=u^2\B_{\mathrm{even}}(s,u-s)^2-s^2\B_{\mathrm{even}}(u,s-u)^2.
\label{eq:even-specialization}
\end{align}
Applying this identity at $(s,s+u)$ and $(s,-u)$ and eliminating their common term shows that
\begin{align}
Q(s,u)\equiv\frac{(s+u)\B_{\mathrm{even}}(s,u)^2-(s-u)\B_{\mathrm{even}}(s,-u)^2}{2u}
\end{align}
satisfies $Q(s,u+s)=Q(s,u)$.
The quotient is regular at $u=0$.
Each homogeneous term is a polynomial invariant under this shift and is therefore independent of $u$.
Evaluating at $u=s$, exchanging $s,u$, and adding gives
\begin{align}
(s+u)\B_{\mathrm{even}}(s,u)^2
=u\B_{\mathrm{diag}}(s)^2+s\B_{\mathrm{diag}}(u)^2,
\qquad \B_{\mathrm{diag}}(s)\equiv\B_{\mathrm{even}}(s,s).
\label{eq:even-diagonal}
\end{align}

The common reflection factor also implies
\begin{align}
\bigl(X_2X_2^\vee+X_3X_3^\vee-X_1X_1^\vee\bigr)^2
=4X_2X_2^\vee X_3X_3^\vee.
\label{eq:reflection-discriminant}
\end{align}
Substituting \eqref{eq:even-diagonal}, setting $(x_1,x_2,x_3)=(s,u,t)$, and cancelling the common factor $[stu(s+t+u)(t+u)^2]^2$, we get
\begin{align}
\bigl[K(u,t)-K(s,t+u)\bigr]^2=0,
\label{eq:KmKeq}
\end{align}
where
\begin{align}
K(s,u)=\frac{\B_{\mathrm{diag}}(s)^2\B_{\mathrm{diag}}(u)^2
-\B_{\mathrm{diag}}(s+u)^2\B_{\mathrm{even}}(s,u)^2}{su}.
\end{align}
The numerator vanishes on both coordinate axes, so $K$ is a regular formal series.
A formal series over the real or complex numbers whose square is zero vanishes.
Setting $t=0$ therefore gives $K(s,u)=K(u,0)$, and symmetry makes $K$ constant.
Using $\B_{\mathrm{even}}(s,u)^2=1-2a_{0,0}su+O((s,u)^4)$ fixes $K=6a_{0,0}$.
Let us write $(\pi\alpha')^2=6a_{0,0}$, which is only a formal parameterization at this stage.
Combining the resulting constant identity with \eqref{eq:even-diagonal} gives
\begin{align}
\B_{\mathrm{even}}(s,u)^2
=\frac{[\B_{\mathrm{diag}}(s)^2+\mathrm i\pi\alpha's]
[\B_{\mathrm{diag}}(u)^2+\mathrm i\pi\alpha'u]}
{\B_{\mathrm{diag}}(s+u)^2+\mathrm i\pi\alpha'(s+u)}.
\label{eq:even-coboundary}
\end{align}
The right-hand side is a multiplicative coboundary, so its normalized square root satisfies the cocycle equation, including the formal $\alpha'=0$ case.
Combining it with the odd cocycle proves
\begin{align}
\boxed{\B(s,u)\B(s+u,t)=\B(u,t)\B(s,t+u),\qquad s,t,u\ \text{independent}.}
\label{eq:splittingCon0}
\end{align}
This is the four-point splitting condition \cite{Arkani-Hamed:2023swr, Cao:2024gln, Arkani-Hamed:2024fyd}, which together with finite spin support at the first massive pole can be used to analytically fix the Veneziano amplitude in positivity bootstrap \cite{Wan:2026pjq}.
Below we will go further and derive the exponential ansatz, which allows the Veneziano amplitude to be bootstrapped without imposing the finite-spin assumption. Appendix~\ref{sec:sym-monodromy} derives the string monodromy condition as another consequence of the master equation, which can likewise serve as the input for uniquely bootstrapping the Veneziano amplitude without the finite-spin assumption.

\paragraph{Exponential representation.}
Taking logarithms of the cocycle equation and differentiating at $t=0$ gives
\begin{align}
\partial_u\log\B(s,u)=\partial_2\log\B(s+u,0)-\partial_2\log\B(u,0).
\end{align}
Defining $F(0)=0$ and $F'(s)=-\partial_2\log\B(s,0)$, and then integrating from $0$ to $u$ yield
\begin{align}
\boxed{\B(s,u)=\exp\bigl[F(s)+F(u)-F(s+u)\bigr].}
\end{align}
The forward expansion fixes this choice of $F$:
\begin{align}
F'(s)=\sum_{n\geq0}a_{n,0}s^{n+1},\qquad
F(s)=\sum_{n\geq0}\frac{a_{n,0}}{n+2}s^{n+2}.
\end{align}
The otherwise arbitrary linear term in $F$ cancels from $\B$.

\paragraph{Fixing the even part.}
Defining $F_{\mathrm{even}}(s)=[F(s)+F(-s)]/2$, $G(s)=s\,\exp[-2F_{\mathrm{even}}(s)]$,
then $\B_{\mathrm{even}}(s,u)^2=suG(s+u)/[(s+u)G(s)G(u)]$.
Substituting this expression into the master equation at $(s,u,-s,-u)$ gives $G(s+u)G(s-u)=G(s)^2-G(u)^2$.
Since $G$ is odd and $G'(0)=1$, twice differentiating in $u$ at $u=0$ yields $G'(s)^2-G(s)G''(s)=1$.
Differentiating in $s$ gives $G'G''-GG'''=0$, so $G''/G$ is constant.
The forward expansion implies $G(s)=s-a_{0,0}s^3+O(s^5)$, and hence
\begin{align}
G''+(\pi\alpha')^2G=0,\qquad G(0)=0,\qquad G'(0)=1.
\end{align}
Therefore
\begin{align}
G(s)=\frac{\sin(\pi\alpha's)}{\pi\alpha'},\qquad F_{\mathrm{even}}(s)=\frac12\log\frac{\pi\alpha's}{\sin(\pi\alpha's)}.
\end{align}
Combining this result with the odd terms in the forward expansion gives
\begin{align}
F(s)=\frac12\log\frac{\pi\alpha's}{\sin(\pi\alpha's)}
+\sum_{k\geq1}\frac{a_{2k-1,0}}{2k+1}s^{2k+1},
\qquad (\pi\alpha')^2=6a_{0,0}.
\end{align}
For $\alpha'=0$, the same differential equation gives $G(s)=s$ and $F_{\mathrm{even}}=0$.

Thus, we give the full ansatz \eqref{eq:full-ansatz}:
\begin{align}
\boxed{\begin{aligned}
\B(s,u)={}&\sqrt{\frac{\pi\alpha' su\sin\!\bigl(\pi\alpha'(s+u)\bigr)}{(s+u)\sin(\pi\alpha's)\sin(\pi\alpha'u)}}\\
&\times\exp\!\left[\sum_{k\geq1}\frac{a_{2k-1,0}}{2k+1}
\bigl(s^{2k+1}+u^{2k+1}-(s+u)^{2k+1}\bigr)\right].
\end{aligned}}
\label{eq:exp_ansatz}
\end{align}
This ansatz, and hence the master equation, imposes stronger constraints than either the splitting condition \eqref{eq:splittingCon0} or the string monodromy condition \eqref{eq:monodromyCon}. 
Its even factor is fixed by $a_{0,0}$, and its remaining freedom consists of the odd forward coefficients.
Conversely, it is straightforward to verify that \eqref{eq:full-ansatz} satisfies the master equation~\eqref{eq:sym-master} for arbitrary formal odd coefficients $a_{2k-1,0}$.
When the exponent converges, the same statement holds as an analytic identity near the origin.
This classifies solutions of the master equation; it does not by itself construct their higher-point completions.

\subsection{Spectral reconstruction and Veneziano uniqueness}
\label{sec:uniqueness}

The exponential solution reconstructs the full four-point amplitude from its regular forward limit.
A forward spectral measure specifies the total exchange strength at each mass, but does not specify its distribution among spins.
We first determine which measures are compatible with the full exponential solution and then examine the spin content of the reconstructed residues.
No finite-spin assumption is imposed in this subsection.

\paragraph{Analytic and dispersive hypotheses.}
We require $\B$ to be jointly analytic near the origin and adopt the positive two-channel dispersive setting of Ref.~\cite{Wan:2026pjq}.
Above the massless multiplet there is a gap $\Lambda^2>0$.
After subtracting the Yang--Mills pole, the forward function has only positive-axis spectral singularities and obeys an exact unsubtracted dispersion relation,
\begin{align}
W(s)&\equiv\lim_{u\to0}\left[A(s,u)+\frac{1}{su}\right]
=\sum_{n\geq0}a_{n,0}s^n
=\int_{\Lambda^2}^{\infty}\frac{\rho(s')}{s'-s}\,\dd s',
\qquad \rho\geq0.
\label{eq:forward-dispersion}
\end{align}
Here $s'$ is a mass-squared variable, and discrete spectral contributions are understood distributionally. The nonnegativity assumption applies to the spectral measure of the stripped two-channel amplitude within the stated bootstrap class. We assume the forward limit and massless subtraction are compatible with this relation at leading weak coupling, without massless loop cuts. The large-contour contribution vanishes, and no independent polynomial or entire subtraction term is allowed. In particular, the same measure determines every $a_{n,0}$, including $a_{0,0}$.

\paragraph{Exponential reconstruction map.}
Define the rescaled positive measure
\begin{align}
\dd\mu(s')\equiv s'\rho(s')\,\dd s',
\qquad
W(s)=\int_{\Lambda^2}^{\infty}\frac{\dd\mu(s')}{s'(s'-s)}.
\label{eq:open-spectral-forward}
\end{align}
The generator in $\B(s,u)=\exp[F(s)+F(u)-F(s+u)]$ satisfies $F'(s)=sW(s)$ and $F(0)=F'(0)=0$.
Integrating gives
\begin{align}
F(q)=\int_{\Lambda^2}^{\infty}\dd\mu(s')\left[-\log\left(1-\frac{q}{s'}\right)-\frac{q}{s'}\right].
\label{eq:open-spectral-generator}
\end{align}
The logarithm is normalized at $q=0$.
The bracket starts at order $q^2/(s')^2$, and $\int(s')^{-2}\dd\mu(s')=a_{0,0}<\infty$ controls convergence near the origin.
The linear subtraction fixes the irrelevant linear freedom in $F$.

\paragraph{A single massive spectral level.}
Consider a measure supported at one mass, $\dd\mu(s')=c\,\delta(s'-s_0)\,\dd s'$, with $s_0\geq\Lambda^2$ and $c>0$.
This does not assume a single exchanged spin.
The reconstruction gives
\begin{align}
&F_{\mathrm{one}}(q)=-c\left[\log\left(1-\frac{q}{s_0}\right)+\frac{q}{s_0}\right],\nonumber\\
&\B_{\mathrm{one}}(s,u)=\left[\frac{s_0(s_0-s-u)}{(s_0-s)(s_0-u)}\right]^c.
\label{eq:open-one-mass-reconstruction}
\end{align}
This is a solution of the cocycle equation, but it must also satisfy the fixed reflection-even part of the full ansatz~\eqref{eq:exp_ansatz}:
\begin{align}
-\frac{c}{2}\log\left(1-\frac{q^2}{s_0^2}\right)=\frac12\log\frac{\pi\alpha' q}{\sin(\pi\alpha' q)}.
\label{eq:open-one-mass-even-test}
\end{align}
For a nonzero positive measure, $a_{0,0}>0$ and we choose $\alpha'=\sqrt{6a_{0,0}}/\pi>0$.
Comparing coefficients in \eqref{eq:open-one-mass-even-test} requires
\begin{align}
\frac{c}{s_0^{2n}}=(\alpha')^{2n}\zeta(2n),\qquad n\geq1.
\end{align}
The equations at $n=1,2$ give $c=\zeta(2)^2/\zeta(4)=5/2$, whereas those at $n=1,3$ give $c^2=\zeta(2)^3/\zeta(6)=35/8$.
They are incompatible.
A nonzero single-mass measure is therefore excluded by the full exponential solution.

\paragraph{Determining the full measure.}
For a general measure, taking the even part of \eqref{eq:open-spectral-generator}, differentiating, and dividing by $q$ gives
\begin{align}
\int_{\Lambda^2}^{\infty}\frac{\dd\mu(s')}{(s')^2-q^2}
=\frac{1-\pi\alpha'q\cot(\pi\alpha'q)}{2q^2}
=\sum_{n=1}^{\infty}\frac{1}{(n/\alpha')^2-q^2}.
\label{eq:open-spectral-transform}
\end{align}
The apparent singularity at $q=0$ is removable.
The last equality is the partial-fraction expansion of the cotangent.

For completeness, this transform determines the measure by the same compact moment argument used in Ref.~\cite{Wan:2026pjq}.
Push forward $(s')^{-2}\dd\mu(s')$ under $x=\Lambda^2/s'$ to obtain a finite positive measure $\dd\hat{\mu}$ on $[0,1]$.
Then
\begin{align}
&\Lambda^{2n}a_{n,0}=\int_0^1x^n\,\dd\hat{\mu}(x), &&n\geq0,
\label{eq:forward-moments}\\
&a_{2m,0}=(\alpha')^{2m+2}\zeta(2m+2), &&m\geq0.
\label{eq:even-forward-zeta}
\end{align}
The bound $\Lambda^{4m}a_{2m,0}\leq a_{0,0}$ implies $\alpha'\Lambda^2\leq1$.
The measure
\begin{align}
\dd\hat{\mu}_{*}(x)=(\alpha')^2\sum_{n=1}^{\infty}\frac{1}{n^2}\delta\left(x-\frac{\alpha'\Lambda^2}{n}\right)\dd x
\end{align}
has precisely these even moments.
Polynomials in $x^2$ are dense in the continuous functions on $[0,1]$, so equality of all even moments implies $\hat{\mu}=\hat{\mu}_{*}$.
Consequently,
\begin{align}
\boxed{\dd\mu(s')=\sum_{n=1}^{\infty}\delta\left(s'-\frac{n}{\alpha'}\right)\dd s'.}
\label{eq:open-reconstructed-measure}
\end{align}
Thus the massive support and its forward weights are derived rather than assumed.
Appendix~\ref{app:forward-moment-reconstruction} records the coefficient-by-coefficient reconstruction of the odd moments.

\paragraph{Reconstructing the amplitude and its spin support.}
Substituting~\eqref{eq:open-reconstructed-measure} into \eqref{eq:open-spectral-generator} gives
\begin{align}
F(q)=\sum_{n=1}^{\infty}\left[-\log\left(1-\frac{\alpha'q}{n}\right)-\frac{\alpha'q}{n}\right]
=\log\Gamma(1-\alpha'q)-\gamma_E\alpha'q.
\end{align}
The linear term cancels from the amplitude, and hence
\begin{align}
\boxed{A(s,u)=-\frac{1}{su}\frac{\Gamma(1-\alpha's)\Gamma(1-\alpha'u)}{\Gamma(1-\alpha'(s+u))}.}
\label{eq:open-reconstructed-amplitude}
\end{align}
This fixes the analytic germ and its analytic continuation under the stated hypotheses.
At the massive pole $s=N/\alpha'$, its positive-propagator pole coefficient is
\begin{align}
\lim_{s\to N/\alpha'}\left(\frac{N}{\alpha'}-s\right)A(s,u)
=\frac{\alpha'}{N!}\prod_{j=1}^{N-1}(j+\alpha'u),
\qquad N=1,2,\ldots.
\label{eq:open-polynomial-residue}
\end{align}
The empty product at $N=1$ is one.
The residue is polynomial in the momentum transfer; restoring the polynomial supersymmetric prefactor of a scalar component preserves polynomiality.
Finite spin support at each massive pole is therefore a consequence of the reconstructed amplitude, not an input to its selection.
The spectrum can still contain unbounded spin as the mass level increases.

If $a_{0,0}=0$, positivity instead forces the finite measure $\hat{\mu}$, and hence all forward coefficients, to vanish.
The exponential then gives $\B=1$ and $A=-1/(su)$, the Yang--Mills limit.

\section{Extension to \texorpdfstring{$\mathcal N=8$}{N=8} supergravity}
\label{sec:gravity}

In the recent work of~\cite{Berman:2026ezk}, from the truncated non-linear constraints a similar exponential solution was conjectured for $\mathcal{N}=8$ supergravity, with a reduced $\SU(4)\times\SU(4)$ R-symmetry. Here we establish that this family exhausts the formal solutions of the gravitational master equation by generalizing our $\mathcal{N}=4$ analysis.
The separate UV assumptions required to select the string amplitude are stated in Section~\ref{sec:sugra-uv}. We first observe that the Grassmann variables of $\mathcal N=8$ on-shell superspace can be divided into two groups of four. Each group can be treated as in the SYM analysis of Section~\ref{sec:sym}, allowing us to reuse many of these identities. We therefore follow the same steps: near the kinematic configuration~\eqref{eq:collision}, we separate the pole terms from the remaining six-point contribution, apply the scalar-parity projection and take the collinear limit to obtain the gravitational master equation.

The momenta and the free parameters $x_1,x_2,x_3$ are exactly the same as in YM.
What changes is the set of allowed factorization channels: without color ordering, all six vanishing three-particle invariants are physical channels. The same configuration also lies on the two-particle factorization channel $s_{36}=0$, whose residue involves a three-point amplitude sewn to a five-point amplitude. As shown below, the three-point amplitudes provide sufficient suppression for the subtracted remainder to stay bounded along a suitable nonsingular approach. Together with the discrete R-symmetry, this makes its scalar-parity projection vanish in the limit.

\subsection{Scalar-parity projection}
\label{sec:sugra-data}

We assume local tree amplitudes whose massless factorization channels exchange only states of the $\mathcal N=8$ supergravity multiplet. The three-point superamplitudes are the standard supergravity ones, while local contact terms compatible with the assumed symmetries are left unrestricted at each EFT order. 

As in the gauge theory, it is convenient to remove the known four-point propagator factors before writing the constraint. In the spinor convention~\eqref{eq:kin}, define
\begin{align}
\mathcal A_4^{\mathrm g}[--++]&=\langle\leg{12}\rangle^4[\leg{34}]^4A^{\mathrm g}(s,t,u),&
A^{\mathrm g}(s,t,u)&=\frac{\kappa^2}{stu}+\widehat A^{\mathrm g}(s,t,u),\nonumber\\
\Bg(s,u)&=\frac{su(-s-u)}{\kappa^2}A^{\mathrm g}(s,-s-u,u),&
\widehat A^{\mathrm g}(s,-s,0)&=\sum_{k\geq0}2^k g_{2k}s^{2k},
\label{eq:sugra-four}
\end{align}
where four-point supersymmetry fixes the state dependence and Bose crossing makes $A^{\mathrm g}(s,t,u)$ fully permutation invariant. The first line strips off the helicity factor while the second introduces the normalized function $\Bg$ and the forward Wilson coefficients $g_{2k}$. Crossing makes the forward series even in $s$, and $\Bg=1+stu\widehat A^{\mathrm g}/\kappa^2$ gives
\begin{align}
\Bg(s,u)=\Bg(u,s),\qquad \Bg(s,0)=1,\qquad \Bg(s,-s)=1.
\end{align}
These regular boundary values of the normalized series will supply the boundary data when we solve the master equation.

We divide the eight flavors of SU(8) into a left $(1,2,3,4)$ and a right group $(5,6,7,8)$, and perform the half-Fourier transform separately in each group. On the six external legs, choose the scalar pattern $(12,13,14,23,34,24)$ in the left group and $(56,57,58,67,78,68)$ in the right group, with one pair from each group on every leg. Each of the eight flavors occurs three times. In \cite{Berman:2026ezk}, the peculiar scalar-parity for supergravity is simply the scalar parity implemented for Yang-Mills, acting on the left and right group simultaneously. Thus we can reuse the identities derived for SYM. We begin by writing, 
\begin{align}
\mathcal A_{6}^{{\mathrm g}{\rm NMHV}}
&=\Delta_6^{\mathrm L}\Delta_6^{\mathrm R}\,\mathcal P_6^{{\mathrm g}{\rm NMHV}},
\nonumber\\
\mathcal P_6^{{\mathrm g}{\rm NMHV}}
&=\sum_{a,b,c,d=1}^{2}\sum_{e,f,h,j=1}^{2}
C_{abef}{}^{cdhj}
(\theta^{a1}\theta^{b2}\hat\theta_{c1'}\hat\theta_{d2'})^{\mathrm L}
(  \theta^{e1}\theta^{f2}\hat\theta_{h1'}\hat\theta_{j2'})^{\mathrm R},
\label{eq:sugra-ward}
\end{align}
where the superscripts $\mathrm L,\mathrm R$ label the two groups of four flavors.
As in \eqref{eq:selected-ward}, this is the selected occupation sector, with $16\times16=256$ independent structures before R-symmetry, not the unrestricted six-point superamplitude.
Other occupations do not enter these scalar projections.
The scalar-parity constraint is
\begin{align}
\mathscr P_{\mathrm g}
&=\overline\varphi^{\mathrm L}\otimes\overline\varphi^{\mathrm R}
-\varphi^{\mathrm L}\otimes\varphi^{\mathrm R},
\qquad
\mathscr P_{\mathrm g}(\mathcal P_6^{\mathrm g})=0.
\label{eq:sugra-parity}
\end{align}
Once again this projection can be evaluated by applying the scalar-projection identity of Section~\ref{sec:scalar-parity} separately to the two flavor groups. 

We can likewise reuse the four-point factorization calculation of Section~\ref{sec:sym}. The internal state sum separates into two four-flavor groups, giving
\begin{align}
\Res_{s_{\leg T}=0}\mathcal A_{6,\mathrm{sel}}^{\mathrm g}
&=\left.
\Delta_6^{\mathrm L}\Delta_6^{\mathrm R}
\mathcal U_{\leg T}^{\mathrm L}\mathcal U_{\leg T}^{\mathrm R}
A^{\mathrm g}_{\leg T}A^{\mathrm g}_{\leg T^c}
\right|_{s_{\leg T}=0}.
\label{eq:sugra-lift}
\end{align}
Here $A^{\mathrm g}_{\leg T}$ and $A^{\mathrm g}_{\leg T^c}$ are the stripped four-point amplitudes on the two sides.

\subsection{Collinear limit and gravitational master equation}
\label{sec:sugra-subtraction}
\label{sec:sugra-collinear}

We now analyze the gravitational amplitude near the same collinear configuration \eqref{eq:collision}. Note that now all $6$ vanishing three-particle Mandelstam variables, 
\begin{align}
\mathcal T
&=\{\leg{123},\leg{234},\leg{345},\leg{134},\leg{135},\leg{146}\}\,,
\label{eq:sugra-channels}
\end{align}
correspond to physical factorization channels. We also have $s_{\leg{36}}=0$, where the amplitude must factorize into a three-point times five-point amplitude. 
Note that in the three point kinematics we are taking both $\langle\leg{36}\rangle$ and $[\leg{36}]$ to vanish. This leads to a vanishing three-particle phase space. As the gravitational three-point amplitude has two powers of momentum, this leads to the factorization contribution to be finite.

For a regular selected contribution, the identities in \eqref{eq:limit-identities} give a kinematic contraction of $C_{abef}{}^{dcjh}-C_{abef}{}^{cdhj}$.
It is therefore sufficient, but not necessary for a single scalar equality, to impose 
\begin{align} 
C_{abef}{}^{cdhj}=C_{abef}{}^{dcjh}. 
\end{align}
We impose this coefficient symmetry through a single discrete R-symmetry of the full amplitude, independently of physical scalar parity.
It is realized by a simultaneous signed flavor exchange in the two non-chiral flavor groups. Suppressing the external-leg label and using the same local flavor labels in each group, we write
\begin{align}
\mathsf R_{\mathrm g}:\quad
(\eta^1,\eta^2,\hat\eta_{1'},\hat\eta_{2'})^X
&\longmapsto
(-\eta^1,\eta^2,\hat\eta_{2'},\hat\eta_{1'})^X,
\qquad X=\mathrm L,\mathrm R.
\end{align}
Writing $\mathsf R_{\mathrm L}$ and $\mathsf R_{\mathrm R}$ for the copies of discrete transformation introduced in Section~\ref{sec:scalar-parity}, acting in the left and right flavor groups respectively, we have $\mathsf R_{\mathrm g}=\mathsf R_{\mathrm L}\mathsf R_{\mathrm R}$. The transformation $\mathsf R_{\mathrm g}$ is an order-two element of $\SU(8)$ and therefore generates a single discrete R-symmetry $\mathbb Z_2^R$. Invariance under $\mathsf R_{\mathrm L}$ or $\mathsf R_{\mathrm R}$ separately is not required. With only the discrete R-symmetry imposed here, the simultaneous scalar-parity condition~\eqref{eq:sugra-parity} is weaker than requiring scalar parity in each flavor group separately, and already suffices for our derivation of the gravitational master equation.

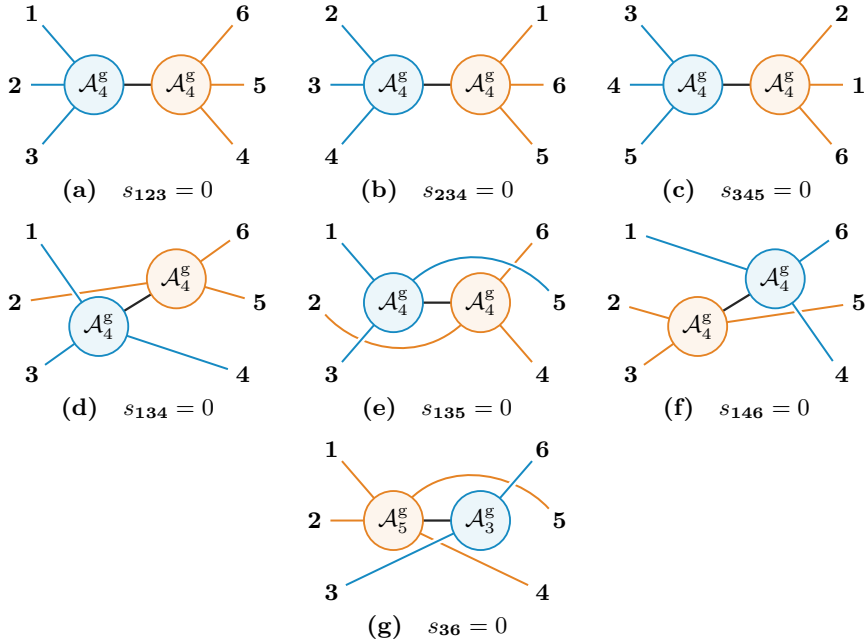
\begin{figure}[!htbp]
\centering
\begingroup
\definecolor{gravitySideA}{HTML}{168AC2}
\definecolor{gravitySideB}{HTML}{E58324}
\begin{tikzpicture}[x=0.9cm,y=0.9cm,
  gravity edge A/.style={draw=gravitySideA,line width=0.75pt,
    shorten >=6pt,preaction={draw=white,line width=2.6pt}},
  gravity edge B/.style={draw=gravitySideB,line width=0.75pt,shorten >=6pt},
  gravity propagator/.style={draw=black!85,line width=0.85pt,
    preaction={draw=white,line width=2.8pt}},
  gravity blob A/.style={circle,draw=gravitySideA,fill=gravitySideA!8,
    line width=0.7pt,minimum size=7.8mm,inner sep=0.8pt,font=\footnotesize},
  gravity blob B/.style={circle,draw=gravitySideB,fill=gravitySideB!8,
    line width=0.7pt,minimum size=7.8mm,inner sep=0.8pt,font=\footnotesize},
  gravity leg/.style={font=\footnotesize,inner sep=1pt,fill=white},
  gravity title/.style={font=\footnotesize,inner sep=0pt}]

\path[use as bounding box] (-2.02,-8.02) rectangle (10.82,1.20);

\newcommand{\gravityPanel}[9]{
  \begin{scope}[shift={#1}]
    \def\gravityLegLayout{1/-1.55/1.05,2/-1.80/0,3/-1.55/-1.05,
      4/1.55/-1.05,5/1.80/0,6/1.55/1.05}
    \ifnum#3=234\relax
      \def\gravityLegLayout{2/-1.55/1.05,3/-1.80/0,4/-1.55/-1.05,
        5/1.55/-1.05,6/1.80/0,1/1.55/1.05}
    \fi
    \ifnum#3=345\relax
      \def\gravityLegLayout{3/-1.55/1.05,4/-1.80/0,5/-1.55/-1.05,
        6/1.55/-1.05,1/1.80/0,2/1.55/1.05}
    \fi
    \foreach \gravityLeg/\gravityX/\gravityY in \gravityLegLayout{
      \coordinate (gravityE\gravityLeg) at (\gravityX,\gravityY);
    }
    \coordinate (gravityA) at #4;
    \coordinate (gravityB) at #5;

    \foreach \gravityLeg in {#7}{
      \ifnum\gravityLeg=2\relax
        \ifnum#3=135\relax
          \draw[gravity edge B] (gravityB)
            .. controls (-0.05,-1.00) and (-1.15,-0.72) .. (gravityE2);
        \else
          \draw[gravity edge B] (gravityB)--(gravityE\gravityLeg);
        \fi
      \else
        \ifnum\gravityLeg=5\relax
          \ifnum#3=36\relax
            \draw[gravity edge B] (gravityB)
              .. controls (-0.05,1.00) and (1.15,0.72) .. (gravityE5);
          \else
            \draw[gravity edge B] (gravityB)--(gravityE\gravityLeg);
          \fi
        \else
          \draw[gravity edge B] (gravityB)--(gravityE\gravityLeg);
        \fi
      \fi
    }

    \foreach \gravityLeg in {#6}{
      \ifnum#3=135\relax
        \ifnum\gravityLeg=5\relax
          \draw[gravity edge A] (gravityA)
            .. controls (0.05,1.00) and (1.15,0.72) .. (gravityE5);
        \else
          \draw[gravity edge A] (gravityA)--(gravityE\gravityLeg);
        \fi
      \else
        \draw[gravity edge A] (gravityA)--(gravityE\gravityLeg);
      \fi
    }
    \draw[gravity propagator] (gravityA)--(gravityB);
    \node[gravity blob A] at (gravityA) {$\mathcal A^{\mathrm g}_{#8}$};
    \node[gravity blob B] at (gravityB) {$\mathcal A^{\mathrm g}_{#9}$};
    \foreach \gravityLeg in {1,...,6}{
      \node[gravity leg] at (gravityE\gravityLeg) {$\mathbf{\gravityLeg}$};
    }
    \node[gravity title] at (0,-1.56)
      {\textbf{(#2)}\quad $s_{\mathbf{#3}}=0$};
  \end{scope}
}

\gravityPanel{(0,0)}{a}{123}{(-0.64,0)}{(0.64,0)}{1,2,3}{4,5,6}{4}{4}
\gravityPanel{(4.40,0)}{b}{234}{(-0.64,0)}{(0.64,0)}{2,3,4}{5,6,1}{4}{4}
\gravityPanel{(8.80,0)}{c}{345}{(-0.64,0)}{(0.64,0)}{3,4,5}{6,1,2}{4}{4}
\gravityPanel{(0,-3.20)}{d}{134}{(-0.57,-0.35)}{(0.57,0.35)}{1,3,4}{2,5,6}{4}{4}
\gravityPanel{(4.40,-3.20)}{e}{135}{(-0.64,0)}{(0.64,0)}{1,3,5}{2,4,6}{4}{4}
\gravityPanel{(8.80,-3.20)}{f}{146}{(0.57,0.35)}{(-0.57,-0.35)}{1,4,6}{2,3,5}{4}{4}
\gravityPanel{(4.40,-6.40)}{g}{36}{(0.64,0)}{(-0.64,0)}{3,6}{1,2,4,5}{3}{5}
\end{tikzpicture}
\endgroup
\caption{The seven channels reaching the collinear configuration~\eqref{eq:collision}. The first row shows the three planar channels used in the gauge-theory proof, with cyclic relabelings of the external legs. The second row shows the three additional three-particle channels in gravity. Every residue in these two rows is a product of four-point superamplitudes and is subtracted in \eqref{eq:sugra-subtraction}; the complementary representatives $\leg{234}\leftrightarrow\leg{156}$ and $\leg{345}\leftrightarrow\leg{126}$ give the set \eqref{eq:sugra-channels}. The final row isolates the $\leg{36}\,|\,\leg{1245}$ channel: a three-point superamplitude sewn to a regular five-point superamplitude, with $p_{\leg3}+p_{\leg6}\to2p_{\leg3}$. Its three-point zeros will make the remainder finite along the approach used in Section~\ref{sec:sugra-collinear}. Each internal line sums over the massless $\mathcal N=8$ multiplet.}
\label{fig:sugra-factorizations}
\end{figure}

We subtract the six three-particle poles before imposing the locus.
For each $T=(abc)$ we may continue its stripped four-point factor as $A^{\mathrm g}(s_{ab},-s_{ab}-s_{bc},s_{bc})$, and similarly for its complement.
These continuations are regular in the local neighborhood apart from the displayed three-particle propagator, agree with the residue on the cut, and preserve the required symmetry.
Thus
\begin{align}
\mathcal P_6^{{\mathrm g},{\rm NMHV}}
&=
\sum_{\leg T\in\mathcal T}
\frac{\mathcal U_{\leg T}^{\mathrm L}\mathcal U_{\leg T}^{\mathrm R}
A^{\mathrm g}_{\leg T}A^{\mathrm g}_{\leg T^c}}{s_{\leg T}}+\mathcal P_{6,\mathrm{rem}}^{{\mathrm g},{\rm NMHV}}\,.
\label{eq:sugra-subtraction}
\end{align}
The remainder inherits supersymmetry and $\mathsf R_{\mathrm g}$-invariance.
It need not be analytic at the coincident-momentum configuration: its only possible local pole is at $s_{\leg{36}}$. Nevertheless, the three-point zeros discussed above keep its coefficients bounded along a nonsingular approach on which both $\langle\leg{36}\rangle$ and $[\leg{36}]$ vanish linearly with nonzero coefficients. Together with $\mathsf R_{\mathrm g}$-invariance and the identities in \eqref{eq:limit-identities}, this ensures that the scalar-parity projection of the remainder tends to zero. The other four three-particle channels and all other two-particle channels remain away from zero at a generic point of the locus.

\paragraph{The projected factorization sum.}
We apply physical scalar parity at nonsingular momenta, then take the specified limit.
The remainder drops out, leaving
\begin{align}
0
&=\sum_{\leg T\in\mathcal T}
\left.
\frac{\mathscr P_{\mathrm g}\bigl(\mathcal U_{\leg T}^{\mathrm L}\mathcal U_{\leg T}^{\mathrm R}\bigr)}
{s_{\leg T}}
\right|_{\mathrm{coll}}
\left.
A_{\leg T}^{\mathrm g}A_{\leg T^c}^{\mathrm g}
\right|_{\mathrm{coll}}.
\label{eq:sugra-projected-poles}
\end{align}
 We now substitute the scalar-projection results from Appendix~\ref{app:sugra-weights} into \eqref{eq:sugra-projected-poles} and express the four-point amplitudes in terms of $\Bg$ using \eqref{eq:sugra-four}. The $\leg{123}$ and $\leg{146}$ contributions vanish. Of the four remaining contributions, the $\leg{234}$ and $\leg{135}$ channels give the same product $\Bg(x_1,x_2)\Bg(x_1+x_2,x_3)$ after applying four-point crossing symmetry. Combining these two terms with the $\leg{345}$ and $\leg{134}$ contributions and cancelling the common nonzero kinematic factor gives the gravitational master equation,
\begin{align}
\boxed{\begin{aligned}
0={}&(x_1-x_2)\Bg(x_1,x_2)\Bg(x_1+x_2,x_3)\\
&+(x_2+x_3)\Bg(x_2,x_3)\Bg(x_1,x_2+x_3)\\
&-(x_1+x_3)\Bg(x_1,x_3)\Bg(x_2,x_1+x_3).
\end{aligned}}
\label{eq:sugra-master}
\end{align}
Like the SYM master equation, this equation yields triangular recursion relations upon inserting the EFT expansion, with the coefficients at each order entering linearly and the nonlinear terms involving only lower-order coefficients. The extension to special kinematic values follows the same argument as in Section~\ref{sec:sym-consistency-equation}. Although the derivation assumes generic shape variables, at each fixed order in the EFT expansion the left-hand side of \eqref{eq:sugra-master} is a polynomial in the independent variables $x_1,x_2,x_3$. Its vanishing on an open set therefore makes it a polynomial identity, which also holds at the special values excluded during the derivation.

\subsection{Exponential ansatz}
\label{sec:sugra-solution}

We now use \eqref{eq:sugra-master} to derive the exponential ansatz. To simplify the calculation, we first exchange $x_1$ and $x_3$ in the master equation and use the symmetry $\Bg(s,t)=\Bg(t,s)$. Subtracting the original equation from the resulting expression and cancelling the common factor $2x_2$ gives the formal power-series identity
\begin{align}
\Bg(x_1,x_2)\Bg(x_1+x_2,x_3)
&=\Bg(x_2,x_3)\Bg(x_1,x_2+x_3).
\label{eq:sugra-product-identity}
\end{align}

To make the consequences of the cocycle equation \eqref{eq:sugra-product-identity} explicit, we take its formal logarithm, differentiate with respect to $x_3$, and then set $x_3=0$. Writing $\partial_2$ for differentiation with respect to the second argument, we obtain
\begin{align}
\partial_2\log\Bg(s,t)
&=\partial_2\log\Bg(s+t,0)-\partial_2\log\Bg(t,0).
\label{eq:sugra-log-derivative}
\end{align}
Thus the derivative at general kinematics is determined entirely by its values at a vanishing second argument. These boundary values are fixed by the regular forward amplitude: using $\Bg(z,0)=1$, the four-point normalization in \eqref{eq:sugra-four}, and crossing symmetry gives
\begin{align}
\partial_2\log\Bg(z,0)
&=-\frac{z^2}{\kappa^2}\widehat A^{\mathrm g}(z,-z,0)
=-\sum_{k\geq0}\frac{2^k g_{2k}}{\kappa^2}z^{2k+2}.
\end{align}

We can now reconstruct the full amplitude by integrating \eqref{eq:sugra-log-derivative} with respect to $t$ at fixed $s$. The boundary condition $\log\Bg(s,0)=0$ fixes the integration constant, giving
\begin{align}
\log\Bg(s,t)
&=\int_0^t\mathrm d\tau\,
\bigl[\partial_2\log\Bg(s+\tau,0)-\partial_2\log\Bg(\tau,0)\bigr].
\end{align}
Substituting the forward expansion and using $u=-s-t$, we obtain the exponential ansatz
\begin{align}
A^{\mathrm g}(s,t,u)
&=\frac{\kappa^2}{stu}
\exp\left[
\sum_{k\geq0}\frac{2^k g_{2k}}{\kappa^2(2k+3)}
\left(s^{2k+3}+t^{2k+3}+u^{2k+3}\right)
\right].
\label{eq:sugra-exponential}
\end{align}
Each term in the exponent has the form $s^N+t^N-(s+t)^N$ and therefore satisfies the logarithmic cocycle identity. Together with crossing symmetry, this makes the three amplitude products in \eqref{eq:sugra-master} equal, so their prefactors cancel. The forward coefficients therefore determine the full solution of the gravitational master equation. This is a simplification compared with SYM, where the corresponding product identity alone does not capture all the constraints.

\subsection{Spectral reconstruction and emergence of infinite spin}
\label{sec:sugra-uv}

The exponential solution~\eqref{eq:sugra-exponential} reconstructs the full amplitude from its regular forward limit, but leaves the forward data undetermined.
We use the dispersive representation of Ref.~\cite{Berman:2026ezk}, following the analytic method of Ref.~\cite{Wan:2026pjq}, to make this remaining freedom explicit.
The exponential is the all-order nonlinear input assumed in that argument and established here; see also Ref.~\cite{Xu:2026kix} for its motivation from KLT.
We first reconstruct an amplitude from a measure supported at one massive level, without specifying the spins there.
Its residue will determine whether the spin expansion terminates.

\paragraph{Dispersive input and reconstruction map.}
We work at leading weak coupling, with a positive gap $\Lambda^2$ above the massless multiplet.
We assume crossing, nonnegative scalar partial-wave spectral weights, and the fixed-transfer sum rules of Ref.~\cite{Berman:2026ezk} for $\mathcal M_{\mathrm{sc}}=s^4A^{\mathrm g}$.
The relevant large-contour terms vanish, and the massless subtraction and forward expansion are assumed to be well defined.
These assumptions give all regular forward coefficients, including $g_0$, without independent subtraction terms.
A sufficient high-energy condition for the contours used there is $\mathcal M_{\mathrm{sc}}/s^3\to0$ at fixed negative transfer.
We additionally assume that, at fixed physical center-of-mass energy
$s>0$ away from isolated $s$-channel poles, the amplitude's boundary
value is analytic as a function of momentum transfer throughout the
interior physical angular interval $-s<t<0$. In particular, no additional
angular branch points are permitted there. This is an explicit assumption
in the massless theory, rather than a consequence of the infrared master
equation. We also require ordinary simple-pole factorization at leading
order. 

Let $\rho_{\mathrm g}(s')=\sum_J\rho_J(s')\geq0$ be the spin-summed density in the normalization of the gravitational forward sum rules, and define
\begin{align}
\dd\mu_{\mathrm g}(s')&\equiv\frac{(s')^3}{\kappa^2}\rho_{\mathrm g}(s')\,\dd s', &g_{2k}&=2^{1-k}\kappa^2\int_{\Lambda^2}^{\infty}\frac{\dd\mu_{\mathrm g}(s')}{(s')^{2k+3}}.
\label{eq:gravity-spectral-moments}
\end{align}
As in the gauge-theory discussion, $s'$ denotes the invariant mass squared.
The sum over $J$ is unrestricted.
Defining $W_{\mathrm g}(s)\equiv\widehat A^{\mathrm g}(s,-s,0)$, we obtain
\begin{align}
W_{\mathrm g}(s)=2\kappa^2\int_{\Lambda^2}^{\infty}\frac{\dd\mu_{\mathrm g}(s')}{s'((s')^2-s^2)}.
\label{eq:gravity-spectral-forward}
\end{align}
The even forward kernel incorporates the crossed channel.
Its vanishing odd Taylor coefficients do not impose additional conditions on the measure.

Write the exponential solution as
\begin{align}
\Bg(s,t)&=\exp[H(s)+H(t)+H(u)],\qquad u=-s-t,\nonumber\\
H(q)&=\sum_{k\geq0}\frac{2^kg_{2k}}{\kappa^2(2k+3)}q^{2k+3},
\qquad H'(q)=\frac{q^2}{\kappa^2}W_{\mathrm g}(q).
\label{eq:gravity-generator-map}
\end{align}
Integrating the spectral representation gives
\begin{align}
H(q)=\int_{\Lambda^2}^{\infty}\dd\mu_{\mathrm g}(s')\left[\log\frac{s'+q}{s'-q}-\frac{2q}{s'}\right].
\label{eq:gravity-spectral-generator}
\end{align}
The bracket begins at order $q^3/(s')^3$, so $\int_{\Lambda^2}^{\infty}(s')^{-3}\dd\mu_{\mathrm g}(s')=g_0/(2\kappa^2)<\infty$ controls the local expansion.
This representation is the logarithmic spectral construction of Ref.~\cite{Berman:2026ezk}, Sec.~6.4.
It automatically gives an odd $H$ with $H'(0)=0$.
Unlike the open generator, it is not subject to an additional fixed reflection-even function.

\paragraph{Reconstruction from one massive level.}
Take the simplest nonzero positive measure,
\begin{align}
\dd\mu_{\mathrm g}(s')=c\,\delta(s'-s_0)\,\dd s',
\qquad c>0,\quad s_0\geq\Lambda^2,
\end{align}
which specifies one mass level.
Substitution into~\eqref{eq:gravity-spectral-generator} gives
\begin{align}
H_{\mathrm{one}}(q)&=c\left[\log\frac{s_0+q}{s_0-q}-\frac{2q}{s_0}\right],\nonumber\\
A^{\mathrm g}_{\mathrm{one}}(s,t,u)&=\frac{\kappa^2}{stu}
\left[\frac{(s_0+s)(s_0+t)(s_0+u)}{(s_0-s)(s_0-t)(s_0-u)}\right]^c.
\label{eq:gravity-one-mass-reconstruction}
\end{align}
The linear terms cancel because $s+t+u=0$. At generic momentum transfer the singularity at $s=s_0$ is $(s_0-s)^{-c}$. Ordinary simple-pole factorization fixes $c=1$, fixing the pole order. Using
\begin{align}
(s_0+s)(s_0+t)(s_0+u)-(s_0-s)(s_0-t)(s_0-u)=2stu,
\end{align}
the reconstructed amplitude becomes
\begin{align}
\boxed{A^{\mathrm g}_{\mathrm{one}}(s,t,u)=\frac{\kappa^2}{stu}+\frac{2\kappa^2}{(s_0-s)(s_0-t)(s_0-u)}.}
\label{eq:ist-amplitude}
\end{align}
This is the infinite spin tower (IST) of Ref.~\cite{Berman:2026ezk}, Sec.~5.3.2, with $s_0=m^2$.
Its rational form and relative normalization have emerged from the forward measure and simple-pole factorization, rather than being inserted as an ansatz.
Its forward coefficients are $g_{2k}=2^{1-k}\kappa^2/s_0^{2k+3}$, so it satisfies the exponential solution and the gravitational master equation to every EFT order.

\paragraph{The reconstructed spin content.}
We now determine which spins must be exchanged at this mass.
For the scalar component, the positive-propagator residue is
\begin{align}
R_{s_0}(t)&\equiv\lim_{s\to s_0}(s_0-s)s^4A^{\mathrm g}_{\mathrm{one}}(s,t,-s-t)
=\frac{2\kappa^2s_0^4}{(s_0-t)(2s_0+t)}\nonumber\\
&=\frac{8\kappa^2s_0^2}{9-z^2},
\qquad z=1+\frac{2t}{s_0}.
\label{eq:ist-residue}
\end{align}
It is regular on the physical angular interval $-1\leq z\leq1$, but is not a polynomial.
Its Legendre expansion therefore cannot terminate.
Positivity of its unbounded even-spin support follows explicitly from
\begin{align}
\frac{1}{9-z^2}&=\sum_{n=0}^{\infty}\frac{z^{2n}}{9^{n+1}}, &z^{2n}&=\sum_{j=0}^{n}\frac{(4j+1)(2n)!}{(2n-2j)!!(2n+2j+1)!!}\,P_{2j}(z).
\label{eq:ist-positive-spin-expansion}
\end{align}
Every displayed coefficient is positive, and the first series converges uniformly on the physical interval.
Hence every even spin has a nonzero positive coefficient in this scalar residue.

\subsection{Finite-spin support and Virasoro--Shapiro uniqueness}
\label{sec:sugra-finite-spin}

To select the string amplitude, we now impose an additional condition on
the spin content of the reconstructed residues. First recall the product
classification of Ref.~\cite{Berman:2026ezk}, Sec.~6.4. Applied to
\eqref{eq:gravity-spectral-generator}, its angular-analyticity argument
localizes the rescaled measure at positive masses with positive integer
weights; simple-pole factorization makes each nonzero weight one. Thus
\begin{align}
A^{\mathrm g}(s,t,u)
&=\frac{\kappa^2}{stu}
\prod_n
\frac{(s_n+s)(s_n+t)(s_n+u)}
     {(s_n-s)(s_n-t)(s_n-u)},
\qquad s_n=m_n^2>0.
\label{eq:sugra-spectral-product}
\end{align}
The masses are distinct and locally finite, with convergence inherited
from the spectral representation. This classification uses the stated
analytic and dispersive hypotheses, not infrared consistency alone.
No finite-spin assumption has entered: the one-mass reconstruction is
already a member of this family, and a unit weight is not a statement
that only one spin is exchanged.

\paragraph{Polynomiality of the first massive residue.}
Let $s_1>0$ be the lowest actual massive pole. We now require only
finitely many spins to have nonzero exchange coefficients there:
\begin{align}
R_1(t)
&\equiv\lim_{s\to s_1}(s_1-s)s^4A^{\mathrm g}(s,t,-s-t)
\nonumber\\
&=\sum_{J=0}^{J_{\max}}r_J
 P_J\left(1+\frac{2t}{s_1}\right),
\qquad J_{\max}<\infty,\quad r_J\geq0.
\label{eq:gravity-finite-spin-condition}
\end{align}
This is the first use of finite-spin support. It makes $R_1(t)$ a
polynomial, rather than merely a function analytic on the physical angular
interval. It neither bounds the number of massive levels nor assumes a
uniform spin bound across the spectrum.

Taking the first residue of~\eqref{eq:sugra-spectral-product} gives
\begin{align}
R_1(t)
=\frac{2\kappa^2s_1^4}{(s_1-t)(2s_1+t)}
\times\prod_{j>1}
\frac{(s_j+s_1)(s_j+t)(s_j-s_1-t)}
     {(s_j-s_1)(s_j-t)(s_j+s_1+t)}.
\label{eq:gravity-first-product-residue}
\end{align}
Polynomiality requires canceling the pole at $t=s_1$. The only possible
numerator cancellation comes from a mass $s_j=2s_1$. The resulting
pole at $t=2s_1$ then requires $3s_1$, and iterating forces every
positive integer multiple of $s_1$ to occur.

Positivity excludes additional offset sequences. For $t>0$, every Legendre
polynomial in~\eqref{eq:gravity-finite-spin-condition} is positive, so the
nonzero finite sum $R_1(t)$ has no positive-$t$ zeros. If $s_*$ were the
smallest mass outside the integer sequence, its numerator in
\eqref{eq:gravity-first-product-residue} would vanish at
$t=s_*-s_1>0$. Canceling that zero would require a pole from the smaller
mass $s_*-s_1$, also outside the integer sequence, a contradiction.
Consequently,
\begin{align}
s_n=ns_1,\qquad n=1,2,\ldots.
\label{eq:gravity-selected-spectrum}
\end{align}
This exhibits how the first-residue condition selects the string spectrum
within the product family.

With $\alpha'=s_1^{-1}$, the spectral moments and reconstructed amplitude
are therefore
\begin{align}
g_{2k}=2^{1-k}\kappa^2(\alpha')^{2k+3}\zeta(2k+3),~~~
\boxed{A^{\mathrm g}(s,t,u)
=\frac{\kappa^2}{stu}
\prod_{q\in\{s,t,u\}}
\frac{\Gamma(1-\alpha'q)}{\Gamma(1+\alpha'q)}~.}
\label{eq:virasoro-shapiro-corollary}
\end{align}
The forward reconstruction fixes the full analytic germ, not merely its
pole locations. This is the conditional Virasoro--Shapiro uniqueness
statement. If the positive massive measure vanishes, all $g_{2k}$ vanish
and the amplitude instead reduces to pure two-derivative supergravity.

The distinction from the gauge theory is now explicit. There the fixed
even generator determines the positive measure and finite-spin residues
follow. Here the exponential leaves the measure free, and a single massive
level already reconstructs the nonpolynomial residue~\eqref{eq:ist-residue}.
In gravity, finite-spin support provides an additional condition for selecting the Virasoro--Shapiro amplitude from the family of exponential solutions.

\subsection{KLT relation}

The master equations for SYM and SUGRA were derived independently from the corresponding six-point factorization. We now relate their solutions by combining two orderings of the gauge-theory four-point function. Given a normalized solution $\B$ of the SYM master equation~\eqref{eq:sym-master}, define
\begin{align}
\Bg(s,t)
&\equiv\frac{\B(s,t)\B(s,u)}{\B(s,-s)},
\qquad u=-s-t.
\label{eq:pairing-gravity-map}
\end{align}
We will show that this construction satisfies the gravitational master equation~\eqref{eq:sugra-master} using the functional identities derived from the SYM master equation.

To verify \eqref{eq:sugra-master}, we will show that its three amplitude products coincide. We first rewrite the definition using the cocycle identity for $\B$ established in Section~\ref{sec:alg-exponential}. Evaluating that identity at $(s,-s,-t)$ and using $\B(0,-t)=1$ gives
\begin{align}
\B(s,u)\B(-s,-t)&=\B(s,-s).
\end{align}
The definition \eqref{eq:pairing-gravity-map} therefore becomes
\begin{align}
\Bg(s,t)
&=\frac{\B(s,t)}{\B(-s,-t)}.
\label{eq:klt-odd-factor}
\end{align}
This form makes the symmetry under $s\leftrightarrow t$ manifest.
The cyclic identity of the odd factor also gives $\Bg(s,t)=\Bg(t,u)=\Bg(u,s)$ on $s+t+u=0$, so the constructed gravitational function is fully crossing symmetric. Dividing the cocycle identity for $\B$ by the same identity with all arguments reversed shows that $\Bg$ also satisfies the cocycle relation. Together with argument symmetry, this yields
\begin{align}
\!\!\Bg(x_1,x_2)\Bg(x_1\!+\!x_2,x_3)
=\Bg(x_2,x_3)\Bg(x_1,x_2\!+\!x_3)
=\Bg(x_1,x_3)\Bg(x_2,x_1\!+\!x_3).
\end{align}
These are precisely the three products appearing in the gravitational master equation~\eqref{eq:sugra-master}. Their common value factors out, leaving the coefficient sum $(x_1-x_2)+(x_2+x_3)-(x_1+x_3)=0$. The gravitational master equation therefore follows.

Finally, we can identify this map with the standard KLT relation by using the opposite-argument value $\B(s,-s)=\pi\alpha's/\sin(\pi\alpha's)$ established in Section~\ref{sec:alg-exponential}, where $(\pi\alpha')^2=6a_{0,0}$. Substituting this result into \eqref{eq:pairing-gravity-map} gives
\begin{align}
\Bg(s,t)
&=\frac{\sin(\pi\alpha's)}{\pi\alpha's}\,
\B(s,t)\B(s,u).
\label{eq:standard-klt}
\end{align}
This is the standard four-point KLT relation in our normalization~\cite{Kawai:1985xq}.
It maps solution families of the two functional equations; it does not assume that an independently specified gravitational EFT is a double copy, nor replace the gravitational remainder analysis.

\section{Summary and discussion}
\label{sec:discussion}

In this work, we have derived quadratic relations for four-point amplitudes in maximally supersymmetric Yang--Mills and supergravity effective field theories from six-point tree-level consistency. These relations lead to triangular recursion relations when expanded in terms of the EFT coefficients. We solve them at the functional level, which allows us to establish the exponential forms of the amplitudes to all orders in the low-energy expansion.
These results give an all-order derivation of structures previously identified through order-by-order calculations in Refs.~\cite{Elvang:2026pmc, Berman:2026ezk}. 

The central mechanism is the elimination of unknown six-point contributions by a scalar-parity projection in a special collinear limit where three (six in gravity) three-particle factorization channels simultaneously go on shell. Supersymmetry and a single discrete R-symmetry transformation make this projection vanish on the regular remainder, while the factorization terms leave finite contributions determined by four-point amplitudes.
Imposing scalar parity on the full amplitude then produces a closed constraint on the four-point data.
In supergravity, the additional two-particle channel is controlled by the known three-point amplitudes.
This formulation highlights the role of scalar parity in the uniqueness proof, and shows that full continuous R-symmetry is not needed for the argument. 

The solutions also clarify the relation between gauge and gravity amplitudes.
The four-point KLT construction maps solutions of the gauge-theory master equation to solutions of the gravitational one: the sine kernel cancels the reflection-even factors, leaving the square of the odd factor.
This relation holds for the solution families before the remaining forward coefficients are fixed by dispersive constraints.

The exponential solutions reveal a useful distinction between the two
spectral reconstruction problems. Under the exact positive forward sum
rules of Section~\ref{sec:uniqueness}, the fixed reflection-even generator
in the gauge theory determines the massive measure and hence the full
Veneziano amplitude. Its polynomial massive residues, and therefore finite
spin support at each mass level, are consequences rather than assumptions.
The gravitational exponential instead reconstructs the amplitude from
forward data without fixing those data. A measure supported at a single
massive level already yields a rational amplitude with a positive,
nonterminating even-spin expansion at that mass. Finite-spin support is
first imposed in Section~\ref{sec:sugra-finite-spin}, where polynomiality of
the lowest massive residue selects the Virasoro--Shapiro spectrum within
the product family obtained under the stated analytic and dispersive
hypotheses. This separates the infrared reconstruction, the emergence of
infinite spin support in an allowed four-point solution, and the additional
spectral condition used for gravitational string uniqueness. None of these
four-point reconstruction statements establishes a higher-point completion
for an arbitrary member of the exponential family.

Several questions remain open. Firstly, one should explore the constraint of maximal supersymmetry for gauge-gravity coupled system. Recent analysis has found non-linear constraints that mixes the single- and double-trace Wilson coefficients in the gluon sector, with promising numeric bounds that isolate Heterotic and IST~\cite{Upcoming}. It will be interesting to find the exact form of identity that gives rise to these non-linear constraints. 
Scalar parity is imposed as a condition on selected six-point components; it would be useful to understand whether it follows from a more general physical principle or can be replaced by a weaker condition.
The reduction of the R-symmetry requirement to a single discrete transformation also motivates a search for analogous projections in theories with less supersymmetry. 

Finally, the master equations are necessary consequences of six-point consistency, but their solutions do not by themselves construct a complete six-point amplitude or establish consistency at higher multiplicity.
Determining which solutions admit such completions would clarify whether further constraints arise beyond those captured by the collinear projection.
Extending the present construction to higher points may provide a systematic way to address this question.\\

\noindent{\bf Acknowledgment}
We would like to thank Fu-Ming Chang and Bi-Sheng Qing for helpful discussions. SYZ acknowledges support from the National Natural Science Foundation of China under grant No.~12475074 and No.~12247103. The research of Y-t H is supported by Taiwan NSTC grant No.~112-2628-M-002-003-MY3. Y-t H acknowledges the use of ChatGPT-6 Astra in facilitating the derivations. ZHW is grateful to Claudia de Rham and Andrew J. Tolley for their hospitality at Imperial College London while this work was in its final stages.

\appendix

\section{Scalar projections on the collinear configuration}
\label{app:scalar-projection-proof}

In this appendix we prove the relations between the minor maps $\Gamma$ and $\widehat\Gamma$ used in Section~\ref{sec:scalar-parity} to compare the two scalar projections in \eqref{eq:two-products}. We work on the collinear configuration \eqref{eq:collision} and use the same little-group representatives as in Section~\ref{sec:scalar-parity}, with $\lambda_{\leg6}=\lambda_{\leg3}$ and $\widetilde\lambda_{\leg6}=\widetilde\lambda_{\leg3}$. For arbitrary coefficient vectors $\widetilde v$ and $v$ satisfying the supersymmetry constraints $\widetilde\lambda\widetilde v=0$ and $\lambda v=0$, we will establish
\begin{align}
\langle\leg{12}\rangle\Gamma_{\leg{456}}(\widetilde v)
&=-\langle\leg{45}\rangle\Gamma_{\leg{123}}(\widetilde v),
~~~~~\langle\leg{25}\rangle\Gamma_{\leg{146}}(\widetilde v)
=\langle\leg{14}\rangle\Gamma_{\leg{235}}(\widetilde v),\nonumber\\
[\leg{35}]\widehat\Gamma_{\leg{136}}(v)
&=[\leg{13}]\widehat\Gamma_{\leg{356}}(v),~~~~~~~~
\langle\leg{13}\rangle\widehat\Gamma_{\leg{124}}(v)
=\langle\leg{35}\rangle\widehat\Gamma_{\leg{245}}(v).
\end{align}
We first derive the two unhatted relations and then the two hatted relations. Finally, we show that their combined proportionality factor satisfies
\begin{equation}
    -\frac{\langle\leg{45}\rangle\langle\leg{25}\rangle\langle\leg{13}\rangle[\leg{35}]}
{\langle\leg{12}\rangle\langle\leg{14}\rangle\langle\leg{35}\rangle[\leg{13}]}
=1,
\end{equation}
completing the kinematic input to the scalar-projection identity. 

\subsection{Unhatted relations}

We begin with the two unhatted relations. The coefficient vector $\widetilde v$ has six components, and the supersymmetry condition $\widetilde\lambda\widetilde v=0$ leaves four independent directions. Two of these directions do not affect the superamplitude, as shifting $\widetilde v_{\leg i}$ by $c_1\lambda_{\leg i}^{1}+c_2\lambda_{\leg i}^{2}$ adds $c_1Q^{1I}+c_2Q^{2I}$ to the Grassmann linear factor, and this additional term is annihilated by the existing factor $\delta^2(Q^I)$. Momentum conservation ensures that these shifts also preserve the supersymmetry condition. We regard coefficient vectors related by these shifts as equivalent, since their Grassmann linear factors give the same expression after multiplication by $\delta^2(Q^I)$. Thus the amplitude, and hence the minor maps $\Gamma$ that extract its scalar components, depends on only two independent degrees of freedom. Two nonzero linear maps on this two-dimensional space must be proportional if they vanish along the same nonzero direction. We will first identify such a direction and then determine the proportionality factor by evaluating the maps on another vector.

We now use the condition $s_{\leg{123}}=0$ to find a common zero direction for $\Gamma_{\leg{123}}$ and $\Gamma_{\leg{456}}$ within these two remaining degrees of freedom. For this purpose, we choose a test vector whose components on legs $\leg4,\leg5,\leg6$ vanish, ensuring that $\Gamma_{\leg{456}}$ vanishes automatically. A convenient choice is $\widetilde\nu_{\leg{123}}=([\leg{23}],[\leg{31}],[\leg{12}],0,0,0)^{\mathsf T}$, which satisfies the supersymmetry condition by the Schouten identity. This is the coefficient vector of $\widehat\Gamma_{\leg{123}}^I(\eta)$ in \eqref{eq:minors}. Substituting it into the minor maps gives
\begin{align}
\Gamma_{\leg{123}}(\widetilde\nu_{\leg{123}})
&=\langle\leg{12}\rangle[\leg{12}]+\langle\leg{23}\rangle[\leg{23}]+\langle\leg{31}\rangle[\leg{31}]
=s_{\leg{123}}=0,\nonumber\\
\Gamma_{\leg{456}}(\widetilde\nu_{\leg{123}})
&=0.
\end{align}
We still have to check that this vector represents a nonzero direction after multiplication by $\delta^2(Q^I)$.
At generic kinematics, the coefficient of $\eta_3^I\eta_4^I\eta_5^I$ in $\delta^2(Q^I)\widehat\Gamma_{123}^I$ is $\langle45\rangle[12]\neq0$.

The two maps therefore vanish on the same nonzero direction and must be proportional. We next determine their proportionality factor.

To determine the proportionality factor, we now use the condition $p_{\leg6}=p_{\leg3}$. With the little-group representatives chosen above, the test vector $w=(0,0,1,0,0,-1)^{\mathsf T}$ satisfies the SUSY constraint $\widetilde\lambda w=\widetilde\lambda_{\leg3}-\widetilde\lambda_{\leg6}=0$. Evaluating the two maps on this vector gives
\begin{align}
\Gamma_{\leg{123}}(w)&=\langle\leg{12}\rangle,&
\Gamma_{\leg{456}}(w)&=-\langle\leg{45}\rangle.
\nonumber
\end{align}
At a generic point where $\langle\leg{12}\rangle\ne0$, these values fix the proportionality factor to be $-\langle\leg{45}\rangle/\langle\leg{12}\rangle$. Thus, for any coefficient vector $\widetilde v$ satisfying the supersymmetry condition, we obtain
\begin{align}
\langle\leg{12}\rangle\Gamma_{\leg{456}}(\widetilde v)
&=-\langle\leg{45}\rangle\Gamma_{\leg{123}}(\widetilde v).
\end{align}

We now apply the same argument to $\Gamma_{\leg{146}}$ and $\Gamma_{\leg{235}}$. At $p_{\leg6}=p_{\leg3}$, momentum conservation and masslessness give $s_{\leg{123}}+s_{\leg{234}}+s_{\leg{146}}=p_{\leg5}^2=0$, so $s_{\leg{146}}=-s_{\leg{123}}-s_{\leg{234}}=0$. The coefficient vector of $\widehat\Gamma_{\leg{146}}^I(\eta)$ in \eqref{eq:minors} therefore provides a common zero direction: its evaluation under $\Gamma_{\leg{146}}$ gives $s_{\leg{146}}$, while $\Gamma_{\leg{235}}$ vanishes because the vector has zero components on legs $\leg2,\leg3,\leg5$. This direction remains nonzero after multiplication by $\delta^2(Q^I)$, as the coefficient of $\eta_{\leg2}^I\eta_{\leg4}^I\eta_{\leg5}^I$ in $\delta^2(Q^I)\widehat\Gamma_{\leg{146}}^I(\eta)$ is $\langle\leg{25}\rangle[\leg{13}]$, which is nonzero at generic kinematics. The two maps are thus proportional. Evaluating them on the same test vector $w$ gives
\begin{align}
\Gamma_{\leg{146}}(w)&=-\langle\leg{14}\rangle,&
\Gamma_{\leg{235}}(w)&=-\langle\leg{25}\rangle.
\nonumber
\end{align}
For $\langle\leg{14}\rangle\ne0$, these values determine the proportionality factor and yield the second unhatted relation:
\begin{align}
\langle\leg{25}\rangle\Gamma_{\leg{146}}(\widetilde v)
&=\langle\leg{14}\rangle\Gamma_{\leg{235}}(\widetilde v).
\end{align}

\subsection{Hatted relations}

We begin with $\widehat\Gamma_{\leg{136}}$ and $\widehat\Gamma_{\leg{356}}$, whose relation follows directly from $\widetilde\lambda_{\leg6}=\widetilde\lambda_{\leg3}$. Substituting this equality into the definitions in \eqref{eq:minors} gives
\begin{align}
\widehat\Gamma_{\leg{136}}(v)
&=[\leg{13}]v_{\leg6}+[\leg{36}]v_{\leg1}+[\leg{61}]v_{\leg3}
=[\leg{13}](v_{\leg6}-v_{\leg3}),\nonumber\\
\widehat\Gamma_{\leg{356}}(v)
&=[\leg{35}]v_{\leg6}+[\leg{56}]v_{\leg3}+[\leg{63}]v_{\leg5}
=[\leg{35}](v_{\leg6}-v_{\leg3}).
\nonumber
\end{align}
Both maps depend on the same combination $v_{\leg6}-v_{\leg3}$, so we immediately obtain
\begin{align}
[\leg{35}]\widehat\Gamma_{\leg{136}}(v)
&=[\leg{13}]\widehat\Gamma_{\leg{356}}(v)
\end{align}

For $\widehat\Gamma_{\leg{124}}$ and $\widehat\Gamma_{\leg{245}}$, we reuse the two-dimensional argument, with $\lambda$ and $\widetilde\lambda$ interchanged and $\delta^2(Q^I)$ replaced by $\delta^2(\hat Q_{I'})$. To find a common zero direction, we choose a test vector supported only on legs $\leg1,\leg3,\leg6$, ensuring that $\widehat\Gamma_{\leg{245}}$ vanishes automatically. A convenient choice is $\nu_{\leg{136}}=(\langle\leg{36}\rangle,0,\langle\leg{61}\rangle,0,0,\langle\leg{13}\rangle)^{\mathsf T}$, the coefficient vector of $\Gamma_{\leg{136},I'}(\hat\eta)$ in \eqref{eq:minors}. Schouten ensures that $\lambda\nu_{\leg{136}}=0$ at general kinematics. Direct substitution gives
\begin{align}
\widehat\Gamma_{\leg{124}}(\nu_{\leg{136}})
&=[\leg{24}]\langle\leg{36}\rangle,\nonumber\\
\widehat\Gamma_{\leg{245}}(\nu_{\leg{136}})
&=0.
\nonumber
\end{align}
On the collinear configuration, $\langle\leg{36}\rangle=0$, so both maps vanish on this vector. To check that it represents a nonzero direction after multiplication by $\delta^2(\hat Q_{I'})$, we examine
\begin{align}
\left.
[\hat\eta_{\leg1 I'}\hat\eta_{\leg2 I'}\hat\eta_{\leg3 I'}]
\bigl(\delta^2(\hat Q_{I'})\Gamma_{\leg{136},I'}(\hat\eta)\bigr)
\right|_{\mathrm{coll}}
&=-\langle\leg{13}\rangle[\leg{12}]\ne0
\end{align}
at generic kinematics on the locus. The two maps therefore annihilate the same nonzero direction in the two-dimensional space and must be proportional. 

To determine the proportionality factor, we choose $\nu_{\leg{123}}=(\langle\leg{23}\rangle,\langle\leg{31}\rangle,\langle\leg{12}\rangle,0,0,0)^{\mathsf T}$. To evaluate the two maps, we first contract momentum conservation with $\lambda_{\leg3}$ and $\widetilde\lambda_{\leg4}$. Using $\langle\leg{36}\rangle=0$ on the collinear configuration gives
\begin{align}
0=\sum_{\leg i}\langle\leg{3i}\rangle[\leg{i4}]
&=\langle\leg{31}\rangle[\leg{14}]
+\langle\leg{32}\rangle[\leg{24}]
+\langle\leg{35}\rangle[\leg{54}].
\nonumber
\end{align}
Substituting the chosen vector into the two maps and using this identity, we obtain
\begin{align}
\widehat\Gamma_{\leg{124}}(\nu_{\leg{123}})
&=[\leg{24}]\langle\leg{23}\rangle+[\leg{41}]\langle\leg{31}\rangle
=-\langle\leg{35}\rangle[\leg{45}],\nonumber\\
\widehat\Gamma_{\leg{245}}(\nu_{\leg{123}})
&=[\leg{45}]\langle\leg{31}\rangle
=-\langle\leg{13}\rangle[\leg{45}].
\nonumber
\end{align}
At generic kinematics where $\langle\leg{13}\rangle[\leg{45}]\ne0$, these values fix the proportionality factor. Thus, for every coefficient vector $v$ satisfying $\lambda v=0$,
\begin{align}
\langle\leg{13}\rangle\widehat\Gamma_{\leg{124}}(v)
&=\langle\leg{35}\rangle\widehat\Gamma_{\leg{245}}(v).
\end{align}

\subsection{Overall normalization}
\label{app:projection-normalization}

It remains to verify that the scalar prefactor produced by the four minor relations equals one. We will establish the following identity at $p_{\leg6}=p_{\leg3}$, keeping the channel invariants explicit:
\begin{align}
\langle\leg{12}\rangle\langle\leg{14}\rangle\langle\leg{35}\rangle[\leg{13}]
+\langle\leg{25}\rangle\langle\leg{45}\rangle\langle\leg{13}\rangle[\leg{35}]
=
\langle\leg{12}\rangle\langle\leg{45}\rangle s_{\leg{234}}
-\langle\leg{15}\rangle\langle\leg{24}\rangle s_{\leg{123}}.
\label{eq:bracket-balance}
\end{align}
The remaining collinear configuration conditions then immediately set the right-hand side to zero and give the required normalization.

To prove this identity, we use momentum conservation to express the relevant square brackets in terms of a common set. Contracting with the spinor pairs $(\lambda_{\leg2},\widetilde\lambda_{\leg3})$, $(\lambda_{\leg1},\widetilde\lambda_{\leg3})$ and $(\lambda_{\leg1},\widetilde\lambda_{\leg4})$ allows us to eliminate $[\leg{13}]$, $[\leg{23}]$ and $[\leg{24}]$ in favor of $[\leg{34}]$, $[\leg{35}]$ and $[\leg{45}]$. We also use Schouten to simplify the accompanying angle-bracket products.

We can now compare the two sides of \eqref{eq:bracket-balance}. Momentum conservation and $p_{\leg6}=p_{\leg3}$ imply $s_{\leg{123}}=s_{\leg{345}}$. Expanding $s_{\leg{234}}=s_{\leg{23}}+s_{\leg{24}}+s_{\leg{34}}$ and $s_{\leg{123}}=s_{\leg{34}}+s_{\leg{35}}+s_{\leg{45}}$, we substitute the momentum-conservation relations into the right-hand side. The terms proportional to $[\leg{45}]$ cancel, and Schouten simplifies the remaining terms. Using $\langle\leg{12}\rangle[\leg{13}]=-\langle\leg{24}\rangle[\leg{34}]-\langle\leg{25}\rangle[\leg{35}]$ together with Schouten then recovers the two bracket products on the left-hand side, completing the proof of \eqref{eq:bracket-balance}.

On the collinear configuration, $s_{\leg{123}}=s_{\leg{234}}=0$, so the two bracket products on the left-hand side of \eqref{eq:bracket-balance} sum to zero. Wherever the denominator is nonzero, the scalar prefactor is therefore
\begin{align}
-\frac{\langle\leg{45}\rangle\langle\leg{25}\rangle\langle\leg{13}\rangle[\leg{35}]}
{\langle\leg{12}\rangle\langle\leg{14}\rangle\langle\leg{35}\rangle[\leg{13}]}
&=1.
\end{align}
The minor relations and \eqref{eq:bracket-balance} contain no bracket denominators. The resulting polynomial scalar-projection identity therefore extends to regular limits of the generic region used in the proof. As an independent check of the normalization, substituting the parameterization \eqref{eq:ambient} with the collinear configuration conditions \eqref{eq:collision-chart} directly into this factor gives unity for generic values of the three free parameters $(\xi_3,\xi_5,\xi_6)$.

\section{Local kinematics and regularity}
\label{app:local-regularity}

This appendix gives an explicit spinor parameterization of the collinear configuration and establishes the local regularity needed in the pole-subtraction argument.
We verify that $x_1,x_2,x_3$ remain locally independent on the collinear locus and that the three planar channel invariants can be included in a local coordinate system in the surrounding kinematic space.

\paragraph{Local spinor coordinates.}
An explicit momentum-conserving chart is given by
\begin{align}
&(\lambda_{\leg1},\lambda_{\leg2},\lambda_{\leg3},\lambda_{\leg4},\lambda_{\leg5},\lambda_{\leg6})
=\begin{pmatrix}
1 & 0 & 1 & 1 & 1 & 1\\
0 & 1 & \xi_1 & \xi_2 & \xi_3 & \xi_4
\end{pmatrix},\nonumber\\
&(\widetilde\lambda_{\leg3},\widetilde\lambda_{\leg4},\widetilde\lambda_{\leg5},\widetilde\lambda_{\leg6})
=\begin{pmatrix}
1 & 0 & \xi_5 & \xi_7\\
0 & 1 & \xi_6 & \xi_8
\end{pmatrix},\nonumber\\
&\widetilde\lambda_{\leg1}
=-\sum_{i=3}^{6}\widetilde\lambda_{\leg i},
\qquad
\widetilde\lambda_{\leg2}=-\sum_{i=3}^{6}\xi_{i-2}\widetilde\lambda_{\leg i}.
\label{eq:ambient}
\end{align}
The kinematic variables used in the main text are
\begin{align}
x_1&=\langle\leg2\leg3\rangle[\leg2\leg3]=-\xi_2-\xi_3\xi_6-\xi_4\xi_8,\nonumber\\
x_2&=\langle\leg3\leg4\rangle[\leg3\leg4]=\xi_2-\xi_1,\nonumber\\
x_3&=\langle\leg5\leg6\rangle[\leg5\leg6]=(\xi_4-\xi_3)(\xi_5\xi_8-\xi_6\xi_7).
\end{align}
The collinear conditions \eqref{eq:collision} are solved by
\begin{align}
\xi_1&=\xi_3(\xi_6-\xi_5),\qquad
\xi_2=\frac{\xi_3(\xi_6-\xi_5)\xi_6}{1+\xi_5},\nonumber\\
\xi_4&=\xi_1,\qquad \xi_7=1,\qquad \xi_8=0.
\label{eq:collision-chart}
\end{align}
Thus $(\xi_3,\xi_5,\xi_6)$ parameterize this locus on the patch $1+\xi_5\ne0$, where
\begin{align}
x_1&=-\frac{\xi_3\xi_6(1+\xi_6)}{1+\xi_5},&x_2&=\frac{\xi_3(\xi_6-\xi_5)(\xi_6-\xi_5-1)}{1+\xi_5},\nonumber\\
x_3&=-\xi_3\xi_6(\xi_6-\xi_5-1),&x_4&=\xi_3(\xi_6-\xi_5)(1+\xi_6)=-x_1-x_2-x_3.
\end{align}

\paragraph{Local independence.}
At $(\xi_3,\xi_5,\xi_6)=(2,-1/2,1)$, the ambient coordinates are $(\xi_1,\ldots,\xi_8)$ $=(3,6,2,3,-1/2,1,1,0)$, and $(x_1,x_2,x_3)=(-8,3,-1)$.
At this point, direct differentiation gives
\begin{align}
\det\frac{\partial(x_1,x_2,x_3)}{\partial(\xi_3,\xi_5,\xi_6)}=-48,
\qquad
\det\frac{\partial(s_{\leg{123}},s_{\leg{234}},s_{\leg{345}})}{\partial(\xi_1,\xi_2,\xi_3)}=\frac32.
\label{eq:chart-ranks}
\end{align}
The first Jacobian is evaluated after restricting to \eqref{eq:collision-chart}; the second is evaluated in the ambient chart \eqref{eq:ambient}, holding the other five coordinates fixed.
The first nonzero determinant shows that the sewn four-point amplitudes probe an open set of three independent kinematic variables even though the three planar channels are simultaneously on shell.
The second shows that these channel invariants can be included in a local coordinate system near the chosen point, as required for the regularity argument below.
Moreover, $\langle\leg{12}\rangle=[\leg{34}]=1$, so both spinor matrices have rank two throughout this chart.

\paragraph{Complete list of channel invariants on the locus.}
For $x_4=-x_1-x_2-x_3$, the fifteen two-particle invariants are
\begin{align}
s_{23}=s_{26}&=x_1,&s_{34}=s_{46}&=x_2,&s_{35}=s_{56}&=x_3,\nonumber\\
s_{13}=s_{16}&=x_4,&s_{12}=-s_{45}&=x_2+x_3,&s_{14}=-s_{25}&=x_1+x_3,\nonumber\\
s_{15}=-s_{24}&=x_1+x_2,&s_{36}&=0.
\label{eq:all-pair-invariants}
\end{align}
The six vanishing $3|3$ partitions are precisely those in \eqref{eq:sugra-channels}; complementary triples are the same channel.
The other four obey
\begin{align}
s_{136}=s_{245}&=2x_4,&s_{236}=s_{145}&=2x_1,\nonumber\\
s_{346}=s_{125}&=2x_2,&s_{356}=s_{124}&=2x_3.
\end{align}
Thus there are six vanishing three-particle denominators and one vanishing two-particle denominator at a generic point, not ten vanishing three-particle denominators.
The generic region excludes zeros of $x_1x_2x_3x_4(x_1+x_2)(x_1+x_3)(x_2+x_3)$ and of the bracket normalizations used in the projection proof.

\paragraph{Regularity after the planar pole subtraction.}
At any fixed EFT order, locality permits each subtracted component to be written over $s_{123}s_{234}s_{345}$, with all nonvanishing factors absorbed into a regular numerator.
Its residue vanishes on each channel away from the other cuts and hence throughout that channel by analyticity.
The independent channel coordinates in \eqref{eq:chart-ranks} therefore divide the numerator one by one; their product divides it as well.
No singularity remains at the intersection.
This argument includes local contact terms and exchange terms regular in this neighborhood.

For gravity, subtracting the six distinct three-particle residues similarly removes their local divisors, even though they need not be independent coordinates.
Near a generic smooth point, each is a distinct nonassociate irreducible factor in the local analytic ring.
If a common numerator is divisible by each, it is divisible by their product.
Only $s_{36}=\langle\leg{36}\rangle[\leg{36}]$ may remain, as discussed in Section~\ref{sec:sugra-collinear}.

\paragraph{Regular supersymmetric coefficients.}
The rank-two spinor matrices have kernels and charge row spaces of constant dimension, so regular local residual bases can be chosen.
For each flavor, complete the two charge coordinates and two residual coordinates by two additional Grassmann coordinates.
This gives a change of variables with a locally nonzero determinant and a regular inverse.
The multiplicative Ward identities factor out the charge product, while the derivative identities remove dependence on the additional coordinates.
Reading off the coefficients of \eqref{eq:selected-ward} therefore introduces no singular denominator.
The same construction applied to both flavor groups proves the gravitational statement.
This extraction is linear over regular kinematic functions and preserves any common zero of all components, as needed for the branchwise three-point zeros in Section~\ref{sec:sugra-collinear}.

\section{Scalar-parity projections of pole terms}
\label{app:pole-projections}

In Section~\ref{sec:scalar-parity}, we discussed how the scalar-parity operator acts on regular contributions. We now turn to the pole terms obtained by sewing two four-point amplitudes. Since these terms are singular at the special kinematics~\eqref{eq:collision}, we first evaluate their scalar-parity projections at generic nearby momenta. The two scalar components have equal residues on the factorization surface $s_{\leg T}=0$, and therefore the difference of their numerators contains a factor of $s_{\leg T}$. Cancelling it against the propagator denominator leaves a finite kinematic coefficient multiplying the product of two four-point amplitudes, which is the contribution occurring in the master equation. We first carry out this calculation for the three planar channels using the numerator $\mathcal U_{\leg T}$, and then reuse the results to determine the gravitational channel weights in SUGRA. 

\subsection{SYM case}

We first evaluate the scalar projections of the pole numerator $\mathcal U_{\leg T}$ at general on-shell momenta using \eqref{eq:two-products}. We begin with the channel $\leg T=\leg{123}$ and write $p_{\leg{123}}=p_{\leg1}+p_{\leg2}+p_{\leg3}$. Substituting the coefficients of $\widehat\Gamma_{\leg{123}}^I(\eta)$ and $\Gamma_{\leg{123},I'}(\hat\eta)$ into the projection formula gives
\begin{align}
\varphi(\mathcal U_{\leg{123}})
&=-s_{\leg{123}}\langle\leg{46}\rangle[\leg{23}]
\langle\leg2|p_{\leg{123}}|\leg6]\langle\leg3|p_{\leg{123}}|\leg4],\nonumber\\
\overline\varphi(\mathcal U_{\leg{123}})
&=0.
\label{eq:sewn-123-projections}
\end{align}
The complementary projection vanishes because its flavor-1 variables occupy legs $\leg4,\leg5,\leg6$, whereas $\widehat\Gamma_{\leg{123}}^1(\eta)$ has support only on legs $\leg1,\leg2,\leg3$. The scalar-parity difference therefore contains an explicit factor of $s_{\leg{123}}$, which cancels the propagator denominator before we take the collinear limit.

Taking the scalar-parity difference in \eqref{eq:sewn-123-projections} and cancelling the factor of $s_{\leg{123}}$ gives
\begin{align}
\frac{\mathscr P(\mathcal U_{\leg{123}})}{s_{\leg{123}}}
&=\langle\leg{46}\rangle[\leg{23}]
\langle\leg2|p_{\leg{123}}|\leg6]\langle\leg3|p_{\leg{123}}|\leg4].
\end{align}
The right-hand side is polynomial in the external spinors, so we can now impose $p_{\leg6}=p_{\leg3}$ directly. The two factors involving leg $\leg6$ depend on its spinors only through the momentum bilinear $|\leg6\rangle[\leg6|=p_{\leg6}$, and hence $\left.\langle\leg{46}\rangle\langle\leg2|p_{\leg{123}}|\leg6]\right|_{p_{\leg6}=p_{\leg3}}=\langle\leg{43}\rangle\langle\leg2|p_{\leg{123}}|\leg3]$.
To simplify the resulting expression, we use $p_{\leg{123}}=p_{\leg1}+p_{\leg2}+p_{\leg3}$ to expand the sandwiches: $\langle\leg2|p_{\leg{123}}|\leg3]
=-\langle\leg{12}\rangle[\leg{13}]$
and $\langle\leg3|p_{\leg{123}}|\leg4]
=-\langle\leg{13}\rangle[\leg{14}]
-\langle\leg{23}\rangle[\leg{24}]$.
The remaining square-bracket product can be reduced using the Schouten identity $[\leg{13}][\leg{24}]
=[\leg{12}][\leg{34}]+[\leg{14}][\leg{23}]$.
We therefore obtain
\begin{align}
\left.\frac{\mathscr P(\mathcal U_{\leg{123}})}{s_{\leg{123}}}\right|_{p_{\leg6}=p_{\leg3}}
&=-\langle\leg{12}\rangle\langle\leg{34}\rangle[\leg{23}]
\bigl(s_{\leg{13}}[\leg{14}]
+\langle\leg{23}\rangle[\leg{13}][\leg{24}]\bigr)\nonumber\\
&=-\langle\leg{12}\rangle\langle\leg{34}\rangle[\leg{23}]
\bigl((s_{\leg{13}}+s_{\leg{23}})[\leg{14}]
+\langle\leg{23}\rangle[\leg{12}][\leg{34}]\bigr)\nonumber\\
&=(s_{\leg{13}}+s_{\leg{23}})
\langle\leg{12}\rangle[\leg{23}]
\langle\leg{34}\rangle[\leg{41}]
-s_{\leg{12}}s_{\leg{23}}s_{\leg{34}}.
\label{eq:sewn-123-partial}
\end{align}
The last equality follows from $[\leg{41}]=-[\leg{14}]$ and $s_{\leg{ij}}=\langle\leg{ij}\rangle[\leg{ij}]$. We can now evaluate this regular expression using the remaining channel conditions in \eqref{eq:collision}.

To simplify \eqref{eq:sewn-123-partial} further on the chosen kinematics~\eqref{eq:collision}, we use the four-spinor identity
\begin{align}
\langle\leg{12}\rangle[\leg{23}]
\langle\leg{34}\rangle[\leg{41}]
&=\frac12\left(
s_{\leg{12}}s_{\leg{34}}
-s_{\leg{13}}s_{\leg{24}}
+s_{\leg{14}}s_{\leg{23}}
-\epsilon
\right),
\end{align}
where our convention is
\begin{align}
\epsilon
&=[\leg{12}]\langle\leg{23}\rangle[\leg{34}]\langle\leg{41}\rangle
-\langle\leg{12}\rangle[\leg{23}]\langle\leg{34}\rangle[\leg{41}].
\label{eq:epsilon}
\end{align}
The channel conditions and momentum conservation give $s_{\leg{13}}=-s_{\leg{12}}-s_{\leg{23}}$, $
s_{\leg{24}}=-s_{\leg{23}}-s_{\leg{34}}$ and $
s_{\leg{14}}=s_{\leg{12}}+s_{\leg{23}}-s_{\leg{34}}$.
Substituting these relations into the four-spinor identity yields $
\left.\langle\leg{12}\rangle[\leg{23}]
\langle\leg{34}\rangle[\leg{41}]\right|_{\mathrm{coll}}
=-s_{\leg{23}}s_{\leg{34}}-\frac{\epsilon}{2}$.
Here and below, $\mathrm{coll}$ denotes evaluation in the collinear limit specified in \eqref{eq:collision}, namely $p_{\leg6}=p_{\leg3}$ and $s_{\leg{123}}=s_{\leg{234}}=s_{\leg{345}}=0$. Equation~\eqref{eq:sewn-123-partial} therefore becomes
\begin{align}
\left.\frac{\mathscr P(\mathcal U_{\leg{123}})}{s_{\leg{123}}}\right|_{\mathrm{coll}}
=-s_{\leg{12}}\left(-s_{\leg{23}}s_{\leg{34}}-\frac{\epsilon}{2}\right)
-s_{\leg{12}}s_{\leg{23}}s_{\leg{34}}=\frac{\epsilon}{2}s_{\leg{12}}
=\frac{\epsilon}{2}(x_2+x_3),
\end{align}
where the last equality uses \eqref{eq:adjacent}.

We now turn to the channels $\leg{234}$ and $\leg{345}$. The calculation follows the same steps, but the two scalar projections are now generally nonzero, so the channel factor $s_{\leg T}$ emerges only after taking their difference. We denote the momentum in channel $\leg T$ by $p_{\leg T}=\sum_{\leg r\in\leg T}p_{\leg r}$. The identity
\begin{align}
\langle\leg i|p_{\leg T}|\leg k]\langle\leg j|p_{\leg T}|\leg l]
-\langle\leg i|p_{\leg T}|\leg l]\langle\leg j|p_{\leg T}|\leg k]
&=s_{\leg T}\langle\leg{ij}\rangle[\leg{kl}]
\label{eq:sewn-sandwich-determinant}
\end{align}
will allow us to make this factor explicit.

For $\leg T=\leg{234}$, direct substitution into \eqref{eq:two-products} gives
\begin{align}
\varphi(\mathcal U_{\leg{234}})
&=s_{\leg{16}}[\leg{23}]\langle\leg{24}\rangle
\langle\leg1|p_{\leg{234}}|\leg4]\langle\leg3|p_{\leg{234}}|\leg1],\nonumber\\
\overline\varphi(\mathcal U_{\leg{234}})
&=s_{\leg{56}}[\leg{23}]\langle\leg{24}\rangle
\langle\leg5|p_{\leg{234}}|\leg4]\langle\leg3|p_{\leg{234}}|\leg5].
\end{align}
Momentum conservation, written as $p_{\leg{234}}=-p_{\leg1}-p_{\leg5}-p_{\leg6}$, implies $\langle\leg5|p_{\leg{234}}|\leg5]
=s_{\leg{234}}-s_{\leg{16}}$ and $
\langle\leg1|p_{\leg{234}}|\leg1]
=s_{\leg{234}}-s_{\leg{56}}$.
Using \eqref{eq:sewn-sandwich-determinant} to pair the sandwiches in each projection, their difference gives
\begin{equation}
\frac{\mathscr P(\mathcal U_{\leg{234}})}{s_{\leg{234}}}
=[\leg{23}]\langle\leg{24}\rangle
\Bigl[
(s_{\leg{56}}-s_{\leg{16}})\langle\leg3|p_{\leg{234}}|\leg4]
+s_{\leg{56}}\langle\leg{53}\rangle[\leg{45}]
-s_{\leg{16}}\langle\leg{13}\rangle[\leg{41}]
\Bigr].
\end{equation}
Thus the channel denominator has again cancelled before any collinear configuration condition is imposed.

For $\leg T=\leg{345}$, the same projection formula gives
\begin{align}
\varphi(\mathcal U_{\leg{345}})
&=s_{\leg{12}}s_{\leg{16}}s_{\leg{35}}s_{\leg{45}},\nonumber\\
\overline\varphi(\mathcal U_{\leg{345}})
&=\langle\leg6|p_{\leg{345}}|\leg3]\langle\leg2|p_{\leg{345}}|\leg4]
\langle\leg3|p_{\leg{345}}|\leg2]\langle\leg4|p_{\leg{345}}|\leg6].
\end{align}
To compare these expressions, we first rearrange the last two sandwiches using \eqref{eq:sewn-sandwich-determinant}:
\begin{align}
\langle\leg3|p_{\leg{345}}|\leg2]\langle\leg4|p_{\leg{345}}|\leg6]
&=\langle\leg3|p_{\leg{345}}|\leg6]\langle\leg4|p_{\leg{345}}|\leg2]
+s_{\leg{345}}\langle\leg{34}\rangle[\leg{26}].
\end{align}
The resulting pairs simplify by the same identity and momentum conservation:
$\langle\leg6|p_{\leg{345}}|\leg3]\langle\leg3|p_{\leg{345}}|\leg6]
=s_{\leg{345}}s_{\leg{123}}-s_{\leg{12}}s_{\leg{45}}$ and $
\langle\leg2|p_{\leg{345}}|\leg4]\langle\leg4|p_{\leg{345}}|\leg2]
=s_{\leg{345}}s_{\leg{146}}-s_{\leg{16}}s_{\leg{35}}$.
Consequently,
\begin{align}
\overline\varphi(\mathcal U_{\leg{345}})
={}&\bigl(s_{\leg{345}}s_{\leg{123}}-s_{\leg{12}}s_{\leg{45}}\bigr)
\bigl(s_{\leg{345}}s_{\leg{146}}-s_{\leg{16}}s_{\leg{35}}\bigr)\nonumber\\&+s_{\leg{345}}\langle\leg{34}\rangle[\leg{26}]
\langle\leg6|p_{\leg{345}}|\leg3]\langle\leg2|p_{\leg{345}}|\leg4].
\end{align}
The term without an explicit factor of $s_{\leg{345}}$ is precisely $\varphi(\mathcal U_{\leg{345}})$. It therefore cancels in the scalar-parity difference, leaving a regular quotient after division by $s_{\leg{345}}$.

We can now impose $p_{\leg6}=p_{\leg3}$ on both quotients. Expanding the sandwiches and using Schouten identities reduces them to the same two four-spinor contractions encountered in the $\leg{123}$ channel:
\begin{align}
\left.\frac{\mathscr P(\mathcal U_{\leg{234}})}{s_{\leg{234}}}\right|_{p_{\leg6}=p_{\leg3}}
={}&s_{\leg{34}}
\langle\leg{12}\rangle[\leg{23}]\langle\leg{34}\rangle[\leg{41}]
-s_{\leg{23}}
[\leg{12}]\langle\leg{23}\rangle[\leg{34}]\langle\leg{41}\rangle\nonumber+s_{\leg{23}}s_{\leg{34}}(s_{\leg{12}}-s_{\leg{14}}),\nonumber\\
\left.\frac{\mathscr P(\mathcal U_{\leg{345}})}{s_{\leg{345}}}\right|_{p_{\leg6}=p_{\leg3}}
={}&(s_{\leg{13}}+s_{\leg{23}})
[\leg{12}]\langle\leg{23}\rangle[\leg{34}]\langle\leg{41}\rangle
-s_{\leg{12}}s_{\leg{23}}s_{\leg{34}}.
\end{align}
On the full collinear configuration, the contraction with angle and square brackets exchanged obeys $\left.[\leg{12}]\langle\leg{23}\rangle
[\leg{34}]\langle\leg{41}\rangle\right|_{\mathrm{coll}}
=-s_{\leg{23}}s_{\leg{34}}+\frac{\epsilon}{2}$, as follows from our definition of $\epsilon$ and the contraction already evaluated above. Using these two contraction identities together with $s_{\leg{12}}-s_{\leg{14}}=s_{\leg{34}}-s_{\leg{23}}$ and $s_{\leg{13}}+s_{\leg{23}}=-s_{\leg{12}}$, the terms containing only Mandelstam invariants cancel. The remaining contributions are
\begin{align}
\left.\frac{\mathscr P(\mathcal U_{\leg{234}})}{s_{\leg{234}}}\right|_{\mathrm{coll}}
&=-\frac{\epsilon}{2}(s_{\leg{23}}+s_{\leg{34}})
=-\frac{\epsilon}{2}(x_1+x_2),\nonumber\\
\left.\frac{\mathscr P(\mathcal U_{\leg{345}})}{s_{\leg{345}}}\right|_{\mathrm{coll}}
&=-\frac{\epsilon}{2}s_{\leg{12}}
=-\frac{\epsilon}{2}(x_2+x_3),
\end{align}
where we have used \eqref{eq:adjacent}.

\subsection{SUGRA case}
\label{app:sugra-weights}

We now apply the same calculation to the gravitational pole terms. For a channel that factorizes into two four-point amplitudes, the two four-flavor groups supply the numerator $\mathcal U_{\leg T}^{\mathrm L}\mathcal U_{\leg T}^{\mathrm R}$. The kinematic scalar projections are identical in the two groups, so simultaneous flavor complementation gives a difference of squares:
\begin{align}
\mathscr P_{\mathrm g}
\bigl(\mathcal U_{\leg T}^{\mathrm L}\mathcal U_{\leg T}^{\mathrm R}\bigr)
=\overline\varphi(\mathcal U_{\leg T})^2
-\varphi(\mathcal U_{\leg T})^2\nonumber=\mathscr P(\mathcal U_{\leg T})
\bigl[\overline\varphi(\mathcal U_{\leg T})
+\varphi(\mathcal U_{\leg T})\bigr].
\end{align}
This factorization allows us to reuse the one-group quotient calculated in the preceding subsection. We can now cancel the channel denominator against the factor of $s_{\leg T}$ in $\mathscr P(\mathcal U_{\leg T})$ and then take the collinear limit. Although the two one-group scalar projections agree on this configuration, their difference divided by $s_{\leg T}$ need not vanish and can leave a finite, nonzero result. The remaining factor, $\overline\varphi(\mathcal U_{\leg T})+\varphi(\mathcal U_{\leg T})$, is regular and can be evaluated directly, giving
\begin{equation}
\left.
\frac{\mathscr P_{\mathrm g}
\bigl(\mathcal U_{\leg T}^{\mathrm L}\mathcal U_{\leg T}^{\mathrm R}\bigr)}
{s_{\leg T}}
\right|_{\mathrm{coll}}
=
2\left.\varphi(\mathcal U_{\leg T})\right|_{\mathrm{coll}}
\left.\frac{\mathscr P(\mathcal U_{\leg T})}{s_{\leg T}}\right|_{\mathrm{coll}}.
\label{eq:sugra-sewn-product}
\end{equation}
Thus each gravitational channel requires two ingredients: the finite scalar-parity quotient and the value of one scalar projection in the collinear limit. We first obtain these quantities for the six channels in \eqref{eq:sugra-channels}, and then combine them with the four-point denominators to determine their contributions to the master equation.

The one-group quotients for the $\leg{123}$, $\leg{234}$, and $\leg{345}$ channels have already been calculated in the preceding subsection. We therefore only need the additional channels $\leg{134}$, $\leg{135}$, and $\leg{146}$. Their calculation follows the same steps: substitute the coefficients of $\widehat\Gamma_{\leg T}^I(\eta)$ and $\Gamma_{\leg T,I'}(\hat\eta)$ into \eqref{eq:two-products}, use momentum conservation and Schouten identities to cancel the channel denominator, and then impose the collinear configuration. This gives
\begin{align}
\left.\frac{\mathscr P(\mathcal U_{\leg{134}})}{s_{\leg{134}}}\right|_{\mathrm{coll}}
&=\frac{\epsilon}{2}(x_1+x_3),&
\left.\frac{\mathscr P(\mathcal U_{\leg{135}})}{s_{\leg{135}}}\right|_{\mathrm{coll}}
&=\frac{\epsilon}{2}(x_1+x_2),\nonumber\\
\left.\frac{\mathscr P(\mathcal U_{\leg{146}})}{s_{\leg{146}}}\right|_{\mathrm{coll}}
&=-\frac{\epsilon}{2}(x_1+x_3).
\end{align}

We next evaluate the extra scalar factor required by \eqref{eq:sugra-sewn-product}. For the three SYM channels, we can reuse the expressions for $\varphi(\mathcal U_{\leg T})$ obtained before taking their scalar-parity differences. The remaining channels are evaluated by the same direct substitution into \eqref{eq:two-products}. In particular, $\varphi(\mathcal U_{\leg{123}})$ and $\varphi(\mathcal U_{\leg{146}})$ contain explicit factors of $s_{\leg{123}}$ and $s_{\leg{146}}$, respectively, and therefore vanish on the chosen configuration. The six scalar values are
\begin{align}
\left.\varphi(\mathcal U_{\leg{123}})\right|_{\mathrm{coll}}
&=\left.\varphi(\mathcal U_{\leg{146}})\right|_{\mathrm{coll}}=0,\nonumber\\
\left.\varphi(\mathcal U_{\leg{234}})\right|_{\mathrm{coll}}
&=-x_1x_3x_4(x_1+x_2),\nonumber\\
\left.\varphi(\mathcal U_{\leg{345}})\right|_{\mathrm{coll}}
&=\left.s_{\leg{12}}s_{\leg{16}}s_{\leg{35}}s_{\leg{45}}\right|_{\mathrm{coll}}
=-x_3x_4(x_2+x_3)^2,\nonumber\\
\left.\varphi(\mathcal U_{\leg{134}})\right|_{\mathrm{coll}}
&=-x_3x_4(x_1+x_3)^2,\nonumber\\
\left.\varphi(\mathcal U_{\leg{135}})\right|_{\mathrm{coll}}
&=-\left.s_{\leg{13}}s_{\leg{15}}s_{\leg{34}}s_{\leg{35}}\right|_{\mathrm{coll}}
=-x_2x_3x_4(x_1+x_2).
\end{align}
Thus the $\leg{123}$ and $\leg{146}$ channels give zero gravitational scalar-parity contributions, even though their one-group quotients are nonzero.

It remains to combine these numerator factors with the two four-point amplitudes. For $\leg T=(\leg a\leg b\leg c)$ and $\leg T^c=(\leg d\leg e\leg f)$, let
\begin{align}
D_{\leg T}
&=s_{\leg{ab}}s_{\leg{bc}}s_{\leg{ca}}
s_{\leg{de}}s_{\leg{ef}}s_{\leg{fd}}.
\end{align}
On the factorization surface, this is the product of the two $stu$ denominators in \eqref{eq:sugra-four}. Here $D_{\leg T}$ without an argument denotes a scalar denominator product, distinct from the minor map $\Gamma_{\leg T}(v)$. Its values in the collinear limit are
\begin{align}
\left.D_{\leg{123}}\right|_{\mathrm{coll}}
=\left.D_{\leg{345}}\right|_{\mathrm{coll}}
&=-x_1x_2x_3x_4(x_2+x_3)^2,
\nonumber\\
\left.D_{\leg{134}}\right|_{\mathrm{coll}}
=\left.D_{\leg{146}}\right|_{\mathrm{coll}}
&=-x_1x_2x_3x_4(x_1+x_3)^2,
\nonumber\\
\left.D_{\leg{234}}\right|_{\mathrm{coll}}
=\left.D_{\leg{135}}\right|_{\mathrm{coll}}
&=-x_1x_2x_3x_4(x_1+x_2)^2.
\end{align}

We denote the corresponding products of normalized four-point functions by $\Pi_{\leg T}^{\mathrm g}$, so that
\begin{align}
\left.A_{\leg T}^{\mathrm g}A_{\leg T^c}^{\mathrm g}\right|_{\mathrm{coll}}
&=\frac{\kappa^4\Pi_{\leg T}^{\mathrm g}}
{\left.D_{\leg T}\right|_{\mathrm{coll}}}.
\end{align}
Four-point crossing symmetry gives
\begin{align}
\Pi_{\leg{123}}^{\mathrm g}
=\Pi_{\leg{345}}^{\mathrm g}
&=\Bg(x_1,x_2+x_3)\Bg(x_2,x_3),
\nonumber\\
\Pi_{\leg{134}}^{\mathrm g}
=\Pi_{\leg{146}}^{\mathrm g}
&=\Bg(x_2,x_1+x_3)\Bg(x_1,x_3),
\nonumber\\
\Pi_{\leg{234}}^{\mathrm g}
=\Pi_{\leg{135}}^{\mathrm g}
&=\Bg(x_1,x_2)\Bg(x_1+x_2,x_3).
\label{eq:sugra-products}
\end{align}

To collect the remaining kinematic factors, define the channel weights
\begin{align}
w_{\leg T}
&=\frac{x_1x_2}{\epsilon}
\left.
\frac{\mathscr P_{\mathrm g}
\bigl(\mathcal U_{\leg T}^{\mathrm L}\mathcal U_{\leg T}^{\mathrm R}\bigr)}
{s_{\leg T}D_{\leg T}}
\right|_{\mathrm{coll}}
=\frac{2x_1x_2}{\epsilon}
\left.
\frac{\varphi(\mathcal U_{\leg T})}{D_{\leg T}}
\frac{\mathscr P(\mathcal U_{\leg T})}{s_{\leg T}}
\right|_{\mathrm{coll}},
\end{align}
where the second equality follows from \eqref{eq:sugra-sewn-product}, and the channel denominator is cancelled before taking the limit. Substituting the scalar projections and denominator products obtained above yields
\begin{align}
\bigl(w_{\leg{123}},w_{\leg{234}},w_{\leg{345}},
w_{\leg{134}},w_{\leg{135}},w_{\leg{146}}\bigr)
&=\bigl(0,-x_1,-x_2-x_3,x_1+x_3,x_2,0\bigr).
\end{align}
This definition extracts the common factor $\kappa^4\epsilon/(x_1x_2)$ from the projected factorization sum, with $\epsilon$ the momentum contraction defined in \eqref{eq:epsilon}. After cancelling this nonzero factor, \eqref{eq:sugra-projected-poles} becomes $\sum_{\leg T\in\mathcal T}w_{\leg T}\Pi_{\leg T}^{\mathrm g}=0$. Using \eqref{eq:sugra-products} then gives the gravitational master equation \eqref{eq:sugra-master}.

\section{String monodromy from master equation}
\label{sec:sym-monodromy}

Here we derive the four-point string monodromy condition from the master relation \eqref{eq:sym-master}. This provides an alternative route to establishing string uniqueness from maximal supersymmetry, since the string monodromy condition has already been shown to uniquely determine the Veneziano amplitude via the analytic bootstrap method of Ref.~\cite{Wan:2026pjq}.

To derive the monodromy condition, we first differentiate the master relation \eqref{eq:sym-master} with respect to $x_3$ and set $(x_1,x_2,x_3)=(s,u,-u)$ to get
\begin{align}
\B(s,u)\B(-s,-u)=\frac{su}{s+u}\,[g(s)+g(u)],
\label{eq:even-product}
\end{align}
where $g(s)=1/s-s[\Ah(s,0)+\Ah(-s,0)]=1/s-2a_{0,0}s+O(s^3)$. In obtaining this relation, we used the normalization $\B(s,0)=1$ and its transverse derivative $\left.\partial_u\B(s,u)\right|_{u=0}=-s\Ah(s,0)$.
We shall then determine the form of $g(s)$, for which we use another consequence of the master equation, \eqref{eq:KmKeq}, where now $K(s,u)=g(s)g(u)-{g(s+u)[g(s)+g(u)]}=6a_{0,0}$.
Expanding $K(s,u)$ at $u=0$ gives a Riccati equation with constant coefficients
\begin{align}
g'(s)+g(s)^2+6a_{0,0}=0 ,
\label{eq:riccati}
\end{align}
involving only the leading four-point Wilson coefficient. Solving this Riccati equation~\eqref{eq:riccati} with $g(s)=1/s+O(s)$ yields $g=G'/G$, where $G(s)=\sin(\pi\alpha's)/(\pi\alpha')$ and $(\pi\alpha')^2=6a_{0,0}$.  Substituting this into \eqref{eq:even-product} leads to:
\begin{align}
\B(s,u)\B(-s,-u)=\frac{s}{G(s)}\frac{u}{G(u)}\frac{G(s+u)}{s+u}.
\label{eq:even-closed}
\end{align}
It remains to relate the different orderings of the physical amplitude, for which $s+t+u=0$. The identity  $\B_{\mathrm{odd}}(s,u)=\B_{\mathrm{odd}}(u,t)$ together with \eqref{eq:even-closed} then implies $[\B(s,u)/\B(u,t)]^2=[sG(t)/(tG(s))]^2$. 
Both ratios have constant term 1, fixing $\B(s,u)/\B(u,t)=sG(t)/(tG(s))$. 
Using $A(s,u)=-\B(s,u)/(su)$ and cycling the variables then gives
\begin{align}
\frac{A(s,u)}{G(t)}=\frac{A(s,t)}{G(u)}=\frac{A(u,t)}{G(s)}.
\end{align}
This together with the identity $G(t)+\mathrm e^{\pm\mathrm i\pi\alpha's}G(u)+\mathrm e^{\mp\mathrm i\pi\alpha'u}G(s)=0$ then leads to the string monodromy condition,
\begin{align}
\boxed{
A(s,u)+\mathrm e^{\pm\mathrm i\pi\alpha's}A(s,t)+\mathrm e^{\mp\mathrm i\pi\alpha'u}A(u,t)=0,\qquad s+t+u=0.
}
\label{eq:monodromyCon}
\end{align}
\section{Reconstructing the odd forward moments}
\label{app:forward-moment-reconstruction}

We record the coefficient form of the positive-measure reconstruction in
Section~\ref{sec:uniqueness}, following Ref.~\cite{Wan:2026pjq}.
The finite positive measure in~\eqref{eq:forward-moments} obeys
\begin{align}
\Lambda^{2n}a_{n,0}
=\int_0^1x^n\,\dd\hat{\mu}(x),
\qquad
a_{2m,0}=(\alpha')^{2m+2}\zeta(2m+2).
\end{align}
For $0\leq x\leq1$, the identity
\begin{align}
x
&=\sum_{k=0}^{\infty}(-1)^k\binom{1/2}{k}(1-x^2)^k
\nonumber\\
&=\lim_{N\to\infty}\sum_{k=0}^{N}\sum_{\ell=0}^{k}
(-1)^{k+\ell}\binom{1/2}{k}\binom{k}{\ell}x^{2\ell}
\end{align}
expresses $x$ as a uniform limit of polynomials in $x^2$.
The absolute binomial coefficients are summable, including at $x=0$;
the partial sums are bounded between zero and one. Thus, for $m\geq0$,
\begin{align}
\Lambda^{4m+2}a_{2m+1,0}
&=\int_0^1x^{2m+1}\,\dd\hat{\mu}(x)
\nonumber\\
&=\lim_{N\to\infty}\sum_{k=0}^{N}\sum_{\ell=0}^{k}
(-1)^{k+\ell}\binom{1/2}{k}\binom{k}{\ell}
\Lambda^{4(m+\ell)}a_{2(m+\ell),0}
\nonumber\\
&=(\alpha')^2\lim_{N\to\infty}
\sum_{k=0}^{N}\sum_{\ell=0}^{k}
(-1)^{k+\ell}\binom{1/2}{k}\binom{k}{\ell}
(\alpha'\Lambda^2)^{2(m+\ell)}\zeta(2m+2\ell+2)
\nonumber\\
&=(\alpha')^2\sum_{j=1}^{\infty}\frac{1}{j^2}
\left(\frac{\alpha'\Lambda^2}{j}\right)^{2m+1}
=\Lambda^{4m+2}(\alpha')^{2m+3}\zeta(2m+3).
\end{align}
The gap bound $\alpha'\Lambda^2\leq1$ established in
Section~\ref{sec:uniqueness} keeps every argument in $[0,1]$.
Uniform convergence against the finite measure and dominated convergence
against the summable $j^{-2}$ weights justify the limiting operations.
Therefore all odd forward coefficients are determined by the even ones.
This is equivalent to the spectral-measure reconstruction
\eqref{eq:open-reconstructed-measure}; it requires no finite-spin assumption.
\bibliographystyle{JHEP}
\bibliography{ref}

\end{document}